\documentclass[a4paper,twocolumn,11pt]{quantumarticle}
\pdfoutput=1
\usepackage[utf8]{inputenc}
\usepackage[english]{babel}
\usepackage[T1]{fontenc}
\usepackage{amsmath}
\usepackage{hyperref}
\usepackage{braket}
\usepackage{multirow}
\usepackage{amsfonts}
\usepackage{amsmath}

\usepackage{tikz}
\usepackage{lipsum}
\usepackage{svg}

\newenvironment{Figure}
  {\par\medskip\noindent\minipage{\linewidth}}
  {\endminipage\par\medskip}

\begin{document}

\title{A hybrid Quantum proposal to deal with 3-SAT problem}

\author{Jose J. Paulet}
\email{paulet@qsimov.com}
\affiliation{Qsimov Quantum Computing S.L, Spain}
\orcid{0000-0002-0777-4119}

\author{Luis F. LLana}
\email{llana@ucm.es}
\affiliation{Complutense university of Madrid, Spain}
\orcid{0000-0003-1962-1504}

\author{Hernan I. de la Cruz}
\email{HernanIndibil.Cruz@uclm.es}
\affiliation{University of Castilla - La Mancha, Spain}
\orcid{0000-0001-6445-5256}

\author{Mauro Mezzini}
\email{mauro.mezzini@uniroma3.it}
\affiliation{Roma Tre University, Italy}
\orcid{0000-0002-5308-0097}

\author{Fernando Cuartero}
\email{Fernando.Cuartero@uclm.es}
\affiliation{University of Castilla - La Mancha, Spain}
\orcid{0000-0001-6285-8860}

\author{Fernando L. Pelayo}
\email{FernandoL.Pelayo@uclm.es}
\affiliation{University of Castilla - La Mancha, Spain}
\orcid{0000-0001-7849-087X}

\maketitle

\begin{abstract}
  Going as far as possible at SAT problem solving is the main aim of our work. For this sake we have  made use of quantum computing from its two, on practice, main models of computation. They have required some reformulations over the former statement of 3-SAT problem in order to accomplish the requirements of both techniques. This paper presents and describes a hybrid quantum computing strategy for solving 3-SAT problems. The performance of this approximation has been tested over a set of representative scenarios when dealing with 3-SAT from the quantum computing perspective.
  
\end{abstract}

\section{Introduction}\label{sec:intro}

The SAT problem is a well-known NP-complete problem \cite{Cook1971TheProcedures} with a wide range of applications in fields such as cryptanalysis \cite{criptoanalysis}, hardware verification \cite{hardware-verification}, AI planning \cite{Kautz1992PlanningSatisfiability.} and medicine ~\cite{Lin2011EfficientATPG}. There are two main approaches to solve the SAT problem from an algorithmic perspective. The first approach is based on the Davis-Putnam-Logemann-Loveland (DPLL) algorithm \cite{Davis1962ATheorem-proving}, which uses backtracking as its core mechanism. The second approach is based on \textit{local search} algorithms, which use heuristics to change from one state to another until reaching a valid interpretation.

Examples of local search algorithms include \textit{hill climbing / gradient descent} and \textit{simulated annealing} \cite{Kirkpatrick1983OptimizationAnnealing}, with the latter being able to leverage the power of annealing quantum computing.

Quantum Annealing, QA, follows a probabilistic algorithm that uses some quantum-mechanical phenomena, such as superposition and tunneling, to search for the global minimum of a given cost function. On practice QA is a powerful optimization technique devoted to deal with some complex problems, most of them combinatorics involving, that are intractable on practice for classical computers. 

Adiabatic quantum computing and quantum annealing go hand in hand, but they do not necessarily have to be the same thing. QA does not necessarily require adiabaticity \cite{adiabatic-no-quantum-annealing}. The time computational complexity of the adiabatic quantum annealing algorithm depends on $T_{a}\cdot max_{s}(||H(s)||)$ \cite{adiabatic-complexity}, where $H(s)$ represents the time-dependent Hamiltonian, and $T_{a}$ is the total annealing time. This ensures that the minimum energy state can be obtained with high probability as long as the calculation takes place in an ideal closed system, perfectly isolated from the interference of environmental energy.

DWave Quantum annealers 
supported by quantum fluctuation, return low energy solutions of an objective function. They are mainly used for optimization problems and probabilistic sampling problems. The first problems search for the lowest energy state and the second problems search for good low energy samples.

D-Wave's quantum computers are protected by enclosures that shield them against electromagnetic interference and keeps the operation temperature below 15 mK. However, this is not enough to eliminate interference, which has the general effect of reducing the probability of ending up in a ground state. Additionally, the annealing time $T_{a}$ is limited to 1 $\mu s$ $\leq T_{a} \leq$ 2000 $\mu s$ which prevent the user from performing in them arbitrarily slow. Therefore, the theoretical guarantees on performance may not apply to these systems. Thus this technique can just be taken as a heuristic solver, which requires empirical approaches for performance analysis \cite{empirical-time-qa}.

There exist some undesirable possible effects on the quality of the solution provided by DWave's quantum annealers as Background Susceptibility,  Flux Noise of the Qubits, DAC Quantization, High-Energy Photon Flux, Readout Fidelity and uncontrolled Programming Errors.

Whatever of these optimization problems to be run within a D-Wave´s Quantum Processing Unit, QPU, requires to build a matrix representing a Quadratic Unconstrained Binary Optimization, QUBO, problem. A QUBO problem is a combinatorial optimization problem defined as:
\begin{equation}
    min\ f(x)=x^{T}Qx
\end{equation}
Where $x$ is a column vector of Boolean variables $(x_{0},\ldots,x_{n-1})$ and $Q$ is a square matrix of size $n \cdot n$ made of the constants of the problem in question.

A more explicit definition of the problem is:
\begin{equation}
min\ f(x)=\sum^{n-1}_{i=0}\sum^{i}_{j=1}q_{ij}x_{i}x_{j}
\end{equation}
Where $x_{i}, x_{j}\in\mathbb{B}$ and $q_{ij}\in\mathbb{R}$ for all $1\leq j\leq i\leq n-1$.

Therefore without loss of generality we just consider upper triangular matrices to represent QUBO problems:

\begin{equation}
\begin{pmatrix}
q_{00} & q_{01} & \cdots & q_{0 n-1}\\
0 & q_{11} & \cdots & q_{1 n-1}\\
\vdots & \vdots & \ddots & \vdots\\
0 & 0 & \cdots & q_{n-1 n-1}
\end{pmatrix}
\end{equation}

These matrices conform the starting point to perform over DWave's quantum annealers.

Quantum computing based on circuit model is another paradigm that leverages the principles of quantum mechanics to perform computations. It performs over quantum circuits made of interconnected quantum gates that manipulate qubits. This conforms a Turing complete computational paradigm.

The most famous search algorithm within this paradigm is Grover's algorithm~\cite{Grover1996ASearch}. This algorithm searches for an element in an unordered space taking a time for it which belongs to \(O(\sqrt{N})\), where $N$ is the size of the search space, i.e., the number of elements among which seeking has to be performed.
It is conformed of two main steps, the oracle operator, which flips the amplitudes of the states corresponding to the solutions, and the inversion about the mean operator, which amplifies the amplitude of those solution states.

The paper is structured as follows, next section provides an state of the art on quantum computing at dealing with SAT problems, section 3 presents the strategies we propose to deal with 3-SAT. Next two sections give a detailed and reproducible description of the setup we have used in order to empirically validate the proposed strategies and the summary of the results obtained. Section conclusion and future work ends the main contribution of this paper. There are 5 appendices devoted to better illustrate fifth section.

\section{Related work}

In \cite{3-sat-dwave} Quantum annealing features to deal with the Boolean Satisfiability Problem (SAT) are shown. There are some many other proposals based on annealing, both simulated\cite{simulated-annealing-local-search-heuristic} and quantum\cite{max-2-sat-108-qubits} in order to solve this problem.




Leporati et al. propose in \cite{3-quantum-algorithms} three different algorithms over quantum computing circuit model to address the 3-SAT problem. These algorithms, in essence, leverage quantum parallelism to compute the 3-SAT function for all possible states simultaneously. Cerf et al. deal with SAT problem as an structured search problem in \cite{Cerf2000NestedProblems}. Focusing on Grover's inspired aproaches we can find a sort of cooperative quantum searching proposed by Cheng et al. in \cite{Cheng2007Quantum3-SAT}. Zhang et al. in \cite{Zhang2020ProcedureAlgorithm} present a hybrid Classical/Quantum perspective to deal with 3-SAT. Yan et al. in \cite{annealing-oracle-grover} present a hybrid quantum approach that build Grover's oracle by quantum annealing.

Paulet et al. propose in \cite{quantum-incremental-sat} a sort of hybrid Classical/Quantum computing approach supported by a heuristics to lead an incremental strategy to perform partial quantum searches over non-structured domains.

\section{Hybrid Quantum approach}

We make use of both quantum annealing as local search approach and, a sort of DPLL circuit based quantum computing powered scheme in order to deal with 3-SAT problem. For the former we have performed computations over DWave´s quantum annealers.

Afterwards, we have reached a quantum state over which Grover´s modified algorithm will finish the search.

More on detail, see fig. \ref{fig:workflow},  QA's output is considered a first approximation from which we reduce the number of superposed states required by Grover´s algorithm.

Quantum annealers to perform over 3-SAT require that the former statement of the problem is transformed into a Max-3SAT problem (a sort of quantitative version of 3-SAT) as an intermediate domain before conforming a QUBO problem which could be sampled over them.

The corresponding output conforms an initial state for Grover's modified algorithm from which the solution is faster reachable with respect to existing algorithms that solve 3-SAT. 

In particular these latter computations, circuit model-based quantum computing, require:
\begin{enumerate}
    \item building the quantum oracle that checks the former 3-SAT formula
    \item superposing some of the qubits
    \item applying Grover´s based algorithm
    \item measuring the registry in order to obtain the desired output.
\end{enumerate}


\begin{widetext}
\begin{Figure}
    \centering
    \includegraphics[width=0.75\linewidth]{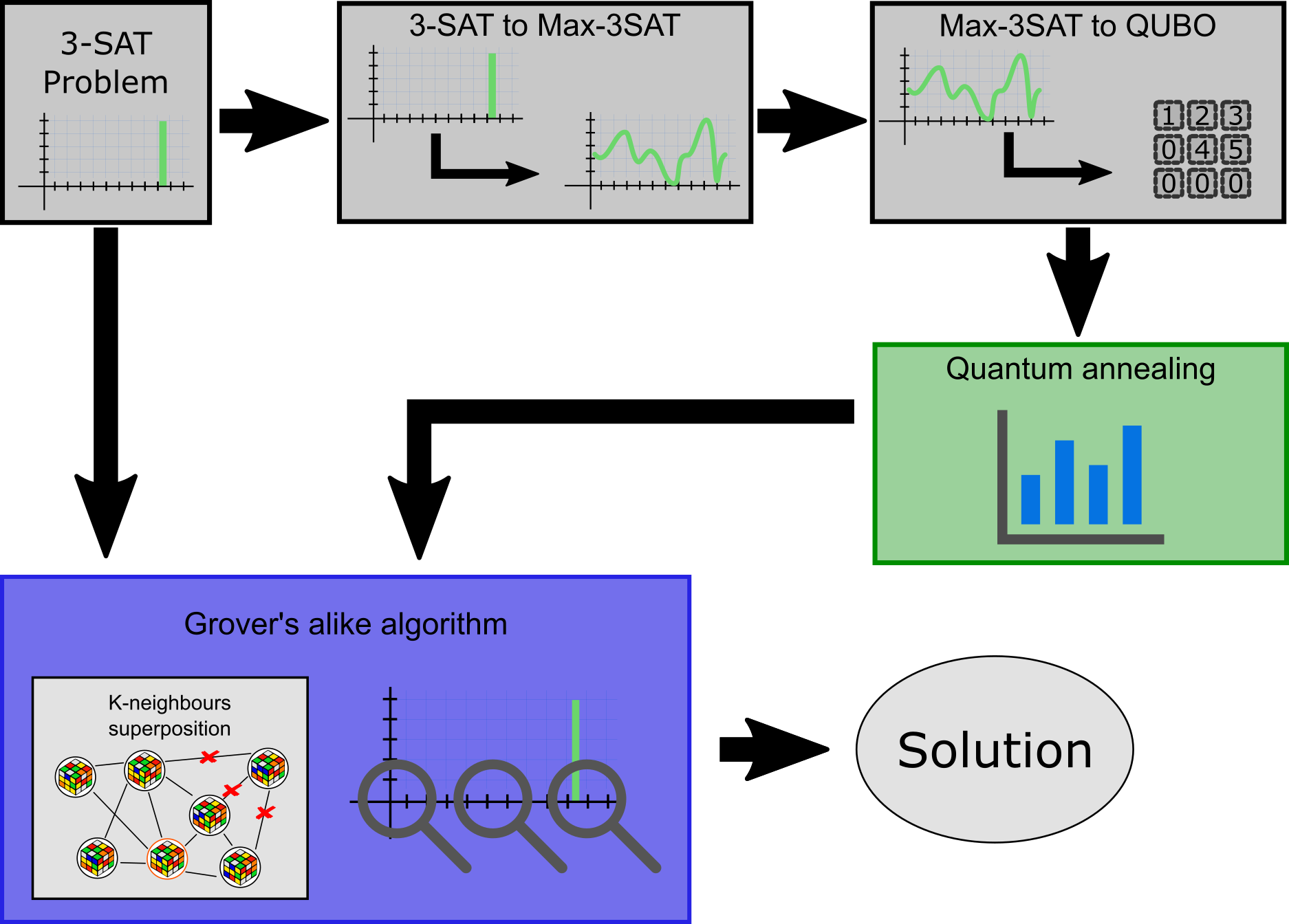}
    \captionof{figure}{Sketch of the hybrid quantum proposal}
    \label{fig:workflow}
\end{Figure}
\end{widetext}

\subsection{First stage - Quantum annealing}
We start using QA as the first approximation to the solution of the problem. In order to do this, a reformulation of the former 3-SAT formula is still required, in particular getting an instance of it as a Quadratic Unconstrained Binary Optimization, QUBO, problem.

Establishing some notation is appropriate in order to describe this process.

Max-SAT\footnote{Unlike the SAT problem, this problem belongs to the class of NP-Hard problems.} is the optimization version of the SAT problem, where the goal is to find the optimal truth assignment based on the maximum number of satisfied clauses, rather than just finding a truth assignment that satisfies a given Boolean formula. We deal with an specific case of weighted Max-SAT (for our case all these weights are set to 1) where weights are assigned to each clause and the objective remains the same: finding the optimal truth assignment. A more restrictive version of Max-SAT is Max-3SAT that, as 3-SAT refers SAT, specifically refers Max-SAT where each clause has exactly 3 literals.

As in Grover's based algorithms, we focus on problems that have a single solution $(|S|=1)$


We start evaluating quantitatively how good a truth assignment is for our 3-SAT, instead of evaluating whether the assignment fulfils the Boolean 3-SAT formula. We move to count the number of clauses satisfied by the truth assignment, namely, treat the 3-SAT problem as Max-3SAT problem. It is immediate that valid truth assignments for a given 3-SAT instance produces in the corresponding Max-3SAT the number of clauses that this 3-SAT instance is made of. 


Let's see in the following example how this is made on practice: Given the Boolean function $f$ corresponding to a 3-SAT problem:
\begin{equation}
\begin{split}
f(&x_{4},x_{3},x_{2},x_{1},x_{0}) = (\overline{x_{0}}\vee \overline{x_{1}}\vee \overline{x_{2}}) \wedge \\
& \wedge (x_{1}\vee \overline{x_{3}}\vee \overline{x_{4}}) \wedge   (x_{4}\vee \overline{x_{2}}\vee \overline{x_{0}})\wedge \\ & \wedge (x_{0}\vee x_{1} \vee \overline{x_{3}}) \wedge (x_{1}\vee x_{2} \vee x_{0}) \\
\end{split}
\end{equation}
The corresponding integer function that defines the Max-3SAT problem is:
\begin{equation}
\begin{split}
    f_{max}(&x_{4},x_{3},x_{2},x_{1},x_{0})= (\overline{x_{0}}\vee \overline{x_{1}}\vee \overline{x_{2}}) + \\
    & + (x_{1}\vee \overline{x_{3}}\vee \overline{x_{4}}) + (x_{4}\vee \overline{x_{2}}\vee \overline{x_{0}}) + \\
    & + (x_{0}\vee x_{1} \vee \overline{x_{3}}) + (x_{1}\vee x_{2} \vee x_{0})
\end{split}
\end{equation}

It is immediate that truth assignment \\ $x_{0}=x_{2}=x_{3}=x_{4}=0$ and $x_{1}=1$ is a model of the 3-SAT problem:
\begin{equation}
\begin{split}
    f(&x_{4},x_{3},x_{2},x_{1},x_{0})= (\overline{0}\vee \overline{1}\vee \overline{0}) \wedge \\
    & (1\vee \overline{0}\vee \overline{0}) \wedge (0\vee \overline{0}\vee \overline{0}) \wedge \\ 
    & \wedge (0\vee 1 \vee \overline{0}) \wedge (1\vee 0 \vee 0) = 1
\end{split}
\end{equation}
And the output on the corresponding Max-3SAT equals the number of clauses:
\begin{equation}
\begin{split}
    f_{max}(&x_{4},x_{3},x_{2},x_{1},x_{0})= (\overline{0}\vee \overline{1}\vee \overline{0}) + \\
    & (1\vee \overline{0}\vee \overline{0}) + (0\vee \overline{0}\vee \overline{0}) + \\
    & + (0\vee 1 \vee \overline{0}) + (1\vee 0 \vee 0) = 5
\end{split}
\end{equation}
On the contrary $x_{0}=x_{1}=x_{2}=x_{3}=x_{4}=0$ is neither a model of the 3-SAT:
\begin{equation}
\begin{split}
    f(&x_{4},x_{3},x_{2},x_{1},x_{0})= (\overline{0}\vee \overline{0}\vee \overline{0}) \wedge \\
    & (0\vee \overline{0}\vee \overline{0}) \wedge (0\vee \overline{0}\vee \overline{0}) \wedge \\
    & (0\vee 0 \vee \overline{0}) \wedge (0\vee 0 \vee 0) = 0
\end{split}
\end{equation}
Nor the output in the corresponding Max-3SAT equals the number of clauses:
\begin{equation}
\begin{split}
    f_{max}(&x_{4},x_{3},x_{2},x_{1},x_{0})= (\overline{0}\vee \overline{0}\vee \overline{0}) + \\
    & (0\vee \overline{0}\vee \overline{0}) + (0\vee \overline{0}\vee \overline{0}) + \\
    & (0\vee 0 \vee \overline{0}) + (0\vee 0 \vee 0) = 4 < 5
\end{split}
\end{equation}
From an instance of a Max-3SAT problem we need to move to an optimization function without restrictions for which we follow the process described in \cite{maxsat-to-qubo} and \cite{penalty-model-max-3-sat}.
Each clause is converted into a function (that will be part of the whole optimization function) which strongly depends on the number of negated literals within the clause as following table shows

\begin{widetext}\hspace*{2cm}
\begin{table}[h]
    \begin{tabular}{|l|l|l|}
    \hline
    Negations & Clause & Equivalent function \\ \hline \hline
    0         &   $(x_{i}\vee x_{j}\vee x_{k})$     &    $x_{i}+x_{j}+x_{k}-x_{i}x_{j}-x_{i}x_{k}-x_{j}x_{k}+x_{i}x_{j}x_{k}$                 \\ \hline
    1         &   $(x_{i}\vee x_{j}\vee \overline{x_{k}})$     &         $1-x_{k}+x_{i}x_{k}+x_{j}x_{k}-x_{i}x_{j}x_{k}$            \\ \hline
    2         &    $(x_{i}\vee \overline{x_{j}}\vee \overline{x_{k}})$    &    $1-x_{j}x_{k}+x_{i}x_{j}x_{k}$                 \\ \hline
    3         &    $(\overline{x_{i}}\vee \overline{x_{j}}\vee \overline{x_{k}})$    &         $1-x_{i}x_{j}x_{k}$            \\ \hline
    \end{tabular}
\end{table}
\end{widetext}


Practically, over our example it produces:
\begin{equation}
\begin{split}
    &f_{max}^{\prime}(x_{4},x_{3},x_{2},x_{1},x_{0})= - x_0 x_1 x_3 - x_0 x_1 + \\
    & + x_0 x_2 x_4- 2 x_0 x_2+ x_0 x_3  + x_0- x_1 x_2  + \\
    & + x_1 x_3 x_4 + x_1 x_3 + x_1  + x_2 - x_3 x_4  - x_3 + 4
\end{split}
\end{equation}

Some cubic terms have been generated at this point. They will be reduced to quadratic by adding some new auxiliary variables together with some penalty functions as stated in \cite{reduction-to-quadratic}, i.e., given a cubic term $x_{i}x_{j}x_{k}$ we have to replace the appearances of $x_{i}x_{j}$ with the variable $y_{ij}$ and afterwards we have to add to the model the penalty function $M(x_{i}x_{j}-2x_{i}y_{ij}-2x_{j}y_{ij}+3y_{ij})$.
This process implies that the number of Boolean variables moves from $n$ in Max-3SAT up to $n+m$ where $m$ refers the number of clauses (which equals the number of cubic terms generated in the previous process \textit{1 per clause at most}).
This is an expensive computational process, specially taken into account that we just want to sample the QUBO problem. 

Therefore, we have decided to follow the integer programming option proposed by Amit et al. in \cite{maxsat-to-qubo-IP}, in advance we take ideas from \cite{maxsat-to-qubo} in order to reduce as much as we can the amount of auxiliary variables.

These latter ideas lead our example to: \begin{equation}
    \begin{split}
        &f_{max}^{\prime\prime}(x_{4},x_{3},x_{2},x_{1},x_{0})=-x_0 x_1 + x_0 x_2 x_4  \\
        &- 2 x_0 x_2+ x_0 x_3 - x_0 y_{13} + x_0 - x_1 x_2    \\
        &+ x_1 x_3+  x_1 + x_2- x_3 x_4 - x_3  + x_4 y_{13} + 4 \\ 
        & - M(x_1 x_3 - 2 x_1 y_{13} - 2 x_3 y_{13} + 3 y_{13})
    \end{split}
\end{equation}

From it, the integer programming option moves our example to: 

\begin{equation}
    \begin{split}
        &f_{QUBO}(x_{4},x_{3},x_{2},x_{1},x_{0})=-x_0 x_1 - 2 x_0 x_2\\
        &  + x_0 x_3 - x_0 y_{13} + x_0 - x_1 x_2 + x_1 x_3 + x_1 \\
        & + x_2- x_3 x_4 - x_3 + x_4 y_{02} + x_4 y_{13} + 4\\
        &- M (x_1 x_3 - 2 x_1 y_{13} - 2 x_3 y_{13} + 3 y_{13}) \\
        &- M (x_0 x_2 - 2 x_0 y_{02} - 2 x_2 y_{02} + 3 y_{02}) 
    \end{split}
\end{equation}




To finish with, proper value for $M$ is computed, according to \cite{maxsat-to-qubo}, as $M=1$. The definite version of $f_{QUBO}$ which coefficients can be directly mapped into DWave QPUs is

\begin{equation}
    \begin{split}
        &f_{QUBO}(x_{4},x_{3},x_{2},x_{1},x_{0})=-x_0 x_1 - 3 x_0 x_2\\
        & + x_0 x_3 + 2 x_0 y_{02}- x_0 y_{13}  - x_1 x_2 + 2 x_1 y_{13} \\
        &+ x_0+ x_1+ 2 x_2 y_{02} + x_2 - x_3 x_4 + 2 x_3 y_{13} \\
        &- x_3+ x_4 y_{02} + x_4 y_{13} - 3 y_{02} - 3 y_{13} + 4
    \end{split}
\end{equation}

\subsection{Second/Last stage - Quantum circuit model}
Once QA has provided its output it is time to describe the quantum circuit-model search algorithms. By this second computation we search for neighbour states based on both Hamming distance and on a sort of cyclical distance. From now on, we will call them \textit{Quantum Hamming search} and \textit{Quantum Cyclical search}.

Henceforth, $S$ denotes the set of solution states of the problem in question and, $\overline{S}$ stands for the non-solution states set of that problem.

\subsubsection{Quantum Hamming search}
It is a quantum computing searching process over Hamming distance, see \cite{Grover1998QuantumTransformation}. It starts on a given quantum state and searches in its Hamming surroundings. Its inputs are:
\begin{itemize}
    \item Number of qubits of the state: $n$
    \item Initial state from which starts computations: $\ket{\gamma}$ (which has been provided by previous QA computing)
    \item Integer value $k$ taken, within Grover´s algorithm, as tentative Hamming distance to solution.
\end{itemize}

Since the correct value for $k$ is not known until finishing the whole search we will denote by $k_f$ the Hamming distance between $\ket{\gamma}$ and the state $\ket{\tau}$ that names the solution of the 3-SAT problem under consideration.

This third parameter $k$ will be increased iteratively from $k=1$ until 
measuring the solution as output of the corresponding Grover's instance. Notice that these values for $k$ hold $k\leq k_f$.




Let's see on detail how does it work:

As usual, an initial superposition is made. In our case the following $H_{\alpha}$ operator is applied to each input qubit:

\begin{equation}
    H_{\alpha}=
        \begin{pmatrix}
        \sqrt{1-\frac{1}{\alpha}}        & \sqrt{\frac{1}{\alpha}} \\
        \sqrt{\frac{1}{\alpha}} & -\sqrt{1-\frac{1}{\alpha}}
    \end{pmatrix}
\end{equation}

Where $\alpha$ is $\displaystyle \frac{n}{k}$ and $k$ is the Hamming distance that we are checking.

This operator performs a non-uniform superposition, that benefits the $k$-nearest states. The resulting probability amplitude of a state at a distance $k$ (either it corresponds to a solution or not) is:
\begin{equation}\label{eq:amp-hamming-k-k}
     amp_{\alpha}=\sqrt{\left(1-\frac{k}{n}\right)^{n-k_{f}}}\cdot\sqrt{\left(\frac{k}{n}\right)^{k_{f}}}
\end{equation}
Afterwards, the amplitude amplification operator is required:
\vspace{-.2cm}
\begin{equation}
    IAM_{\alpha}=2\cdot H_{\alpha}^{\otimes n}\ket{\gamma}\bra{\gamma}H_{\alpha}^{\dagger\otimes n}-I_{n}
\end{equation}

In a nutshell, the corresponding quantum circuit (once a non-measurable -1 global phase has been added) is that depicted in fig. \ref{fig:iam-hamming}:
\begin{widetext}
\begin{Figure}
    \centering
    \includegraphics[width=0.35\linewidth]{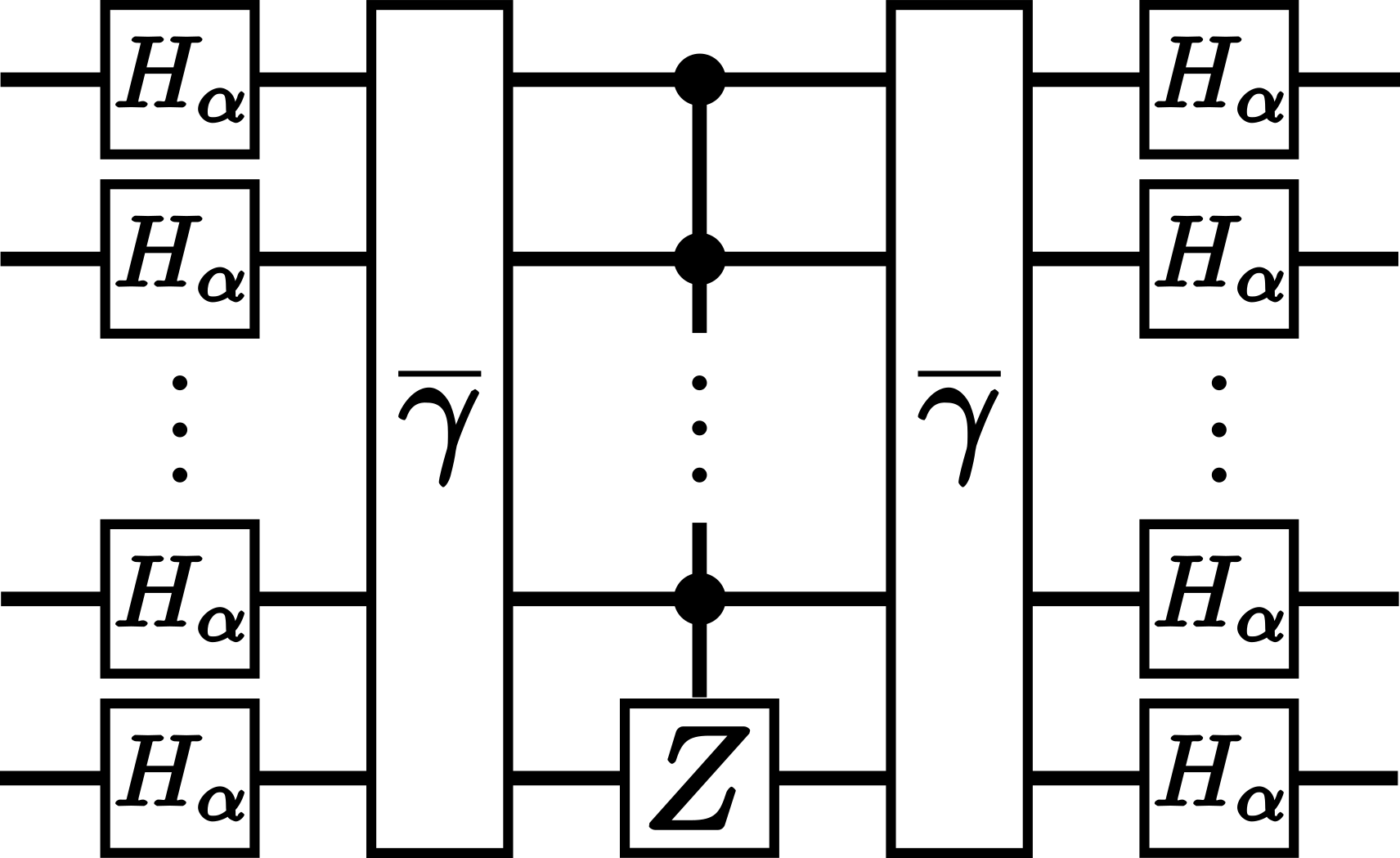}
    \captionof{figure}{Inversion about the mean (Hamming distance)}
    \label{fig:iam-hamming}
\end{Figure}
\end{widetext}

 Our algorithm executes $t_{\alpha}$ times the Grover's iterative operator $G_{\alpha}$ defined as follows:
\begin{equation}
    G_{\alpha}=(2\cdot H_{\alpha}^{\otimes n}\ket{\gamma}\bra{\gamma}H_{\alpha}^{\dagger\otimes n}-I_{n})\cdot U_{f}
\end{equation}
Where $U_{f}$ refers, as usual, the quantum oracle that just inverts the sign of the solution state.


Regarding the initial superposition, let us notice that the angle between the no solution states $\ket{\overline{S}}$ and the initial superposition $\ket{\psi_{\alpha}}$ is $\theta_{\alpha}$ as it is shown in fig \ref{fig:alpha-geometry}.
 \begin{widetext}
\begin{Figure}
    \centering
    \includegraphics[width=0.45\linewidth]{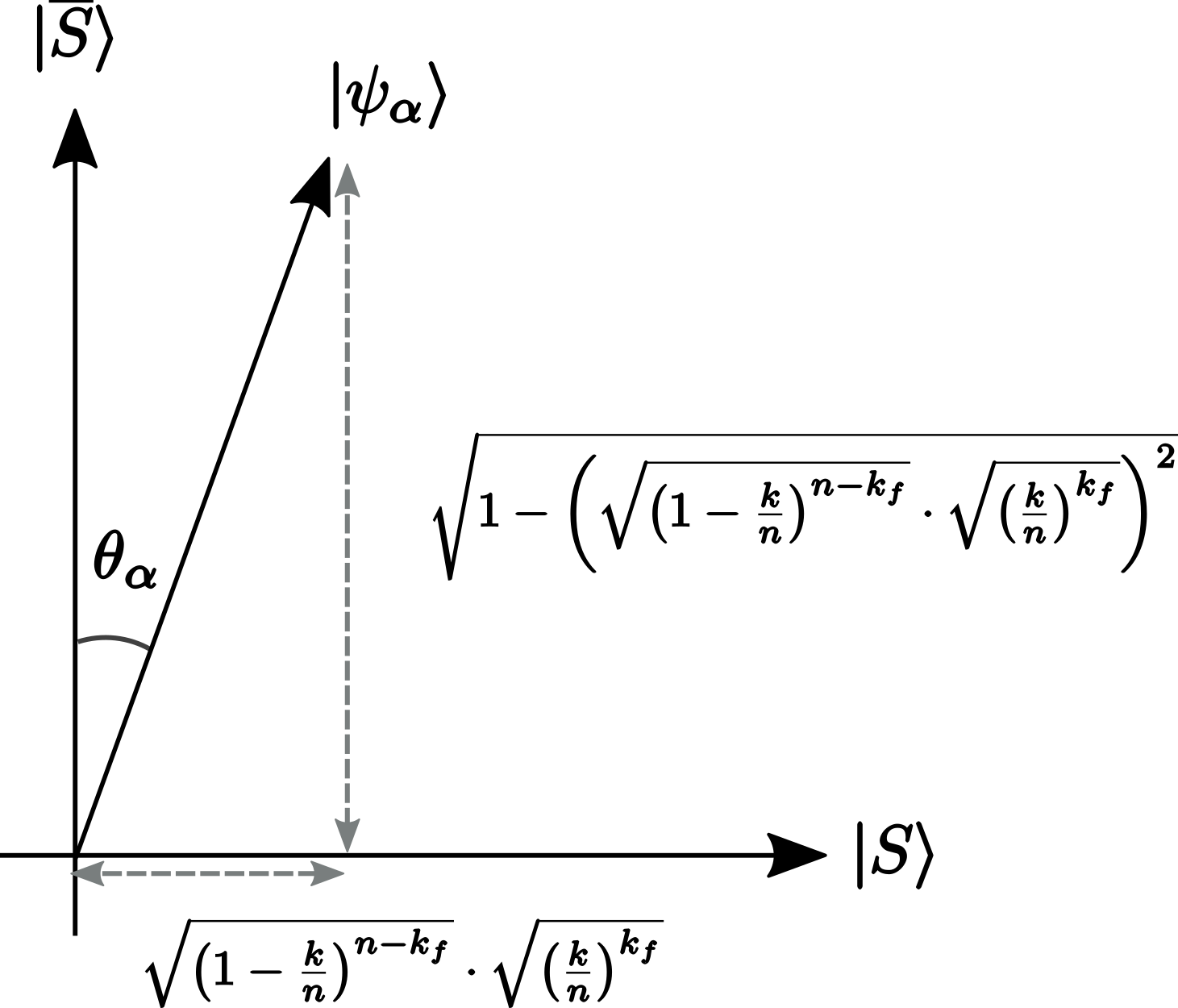}
    \captionof{figure}{Graphical representation of the initial superposition (Quantum Hamming search)}
    \label{fig:alpha-geometry}
\end{Figure}
\end{widetext}
Therefore:
\vspace{-.015cm}
\begin{equation}
\theta_{\alpha}=\sin^{-1}\left(\sqrt{\left(1-\frac{k}{n}\right)^{n-k_{f}}}\cdot\sqrt{\left(\frac{k}{n}\right)^{k_{f}}}\right)
 \end{equation}
 
It is straightforwarded that each iteration involves a rotation of $2\cdot\theta_{\alpha}$, from which the probability amplitude of a solution state after $t_{\alpha}$ iterations will be:

 \begin{widetext}
 \begin{equation}
    amp_{t_{\alpha}}=\sin\left(\theta_{\alpha}\cdot\left(2\cdot t_{\alpha}+1\right)\right)=\sin\left(\sin^{-1}\left(\sqrt{\left(1-\frac{k}{n}\right)^{n-k_{f}}}\cdot\sqrt{\left(\frac{k}{n}\right)^{k_{f}}}\right)\cdot\left(2\cdot t_{\alpha}+1\right)\right)
 \end{equation}
 \end{widetext}
 
In order to calculate the number of iterations required to measure the solution state with a probability acceptably close to 1, we force the following condition to be met:

 \begin{equation}
     \theta_{\alpha}\left(2\cdot t_{\alpha}+1\right)=\frac{\pi}{2}
 \end{equation}
 
 From which we can compute $t_{\alpha}$:
 
 \begin{equation}\label{eq:iterations-hamming}
     t_{\alpha}=\frac{\pi}{4\cdot\sin^{-1}\left(\sqrt{\left(1-\frac{k}{n}\right)^{n-k_{f}}}\cdot\sqrt{\left(\frac{k}{n}\right)^{k_{f}}}\right)}-\frac{1}{2}
 \end{equation}
As previously stated, value of $k_f$ is not known in advance for which we assume within each iteration that the Hamming distance we are considering is the right one, which implies $k_f=k$ and consequently with this assumption we compute  the number of iterations to be performed at this time as $Round (t_{\alpha})$, i.e., $\lfloor t_{\alpha}\rceil$.

These $\lfloor t_{\alpha}\rceil$ are the values to be added per each value of $k$,in order to get the key element to be compared.

\subsubsection{Quantum Cyclical search}\label{subsec:quantum-cyclic-search}
The original Grover's algorithm uses the Walsh-Hadamard transformation to search all along the problem domain.

Some other quantum searching algorithms, see \cite{Grover1998QuantumTransformation, grover-generalized-initial-amplitude}, do replace the Walsh-Hadamard transformation within this task.

In particular, we support our proposal on a replacement of this transformation by a kind of  neighbourhood search. In particular this neighbourhood is defined over a cyclical distance which will be defined in subsection \ref{QCS}. 

This algorithm searches among the $2^{r}$ closest states to the state $\ket{\gamma}$ provided in first stage, i.e., it first searches in the states $[(\gamma-2^{r-1}+1)\mod{2^{n}},(\gamma+2^{r-1})\mod{2^{n}}]$ for the first iteration (corresponding to the states which distance belongs to $[0,2^{r-1}]$), then it searches in the next $2^{r}$ closest states to $\ket{\gamma}$ avoiding previously searched states (corresponding to the states having a distance belonging to $(2^{r-1},2^r]$), and so on.

As a matter of example, the distance between state $0$ and state $2^{n}-1$ is just $1$ as a consequence of our cyclical assumption $mod\ 2^n$.

Let us describe, in an algorithmic way, how the corresponding superposition on the quantum states could be generated.

We have named \textit{Range splitter operator} $\left(U_{r}\right)$, which quantum circuit representation is shown in fig. \ref{fig:range-splitter-operator}, the piece of quantum code responsible for this task.
\begin{widetext}
\begin{Figure}
    \centering
    \includegraphics[width=0.45\linewidth]{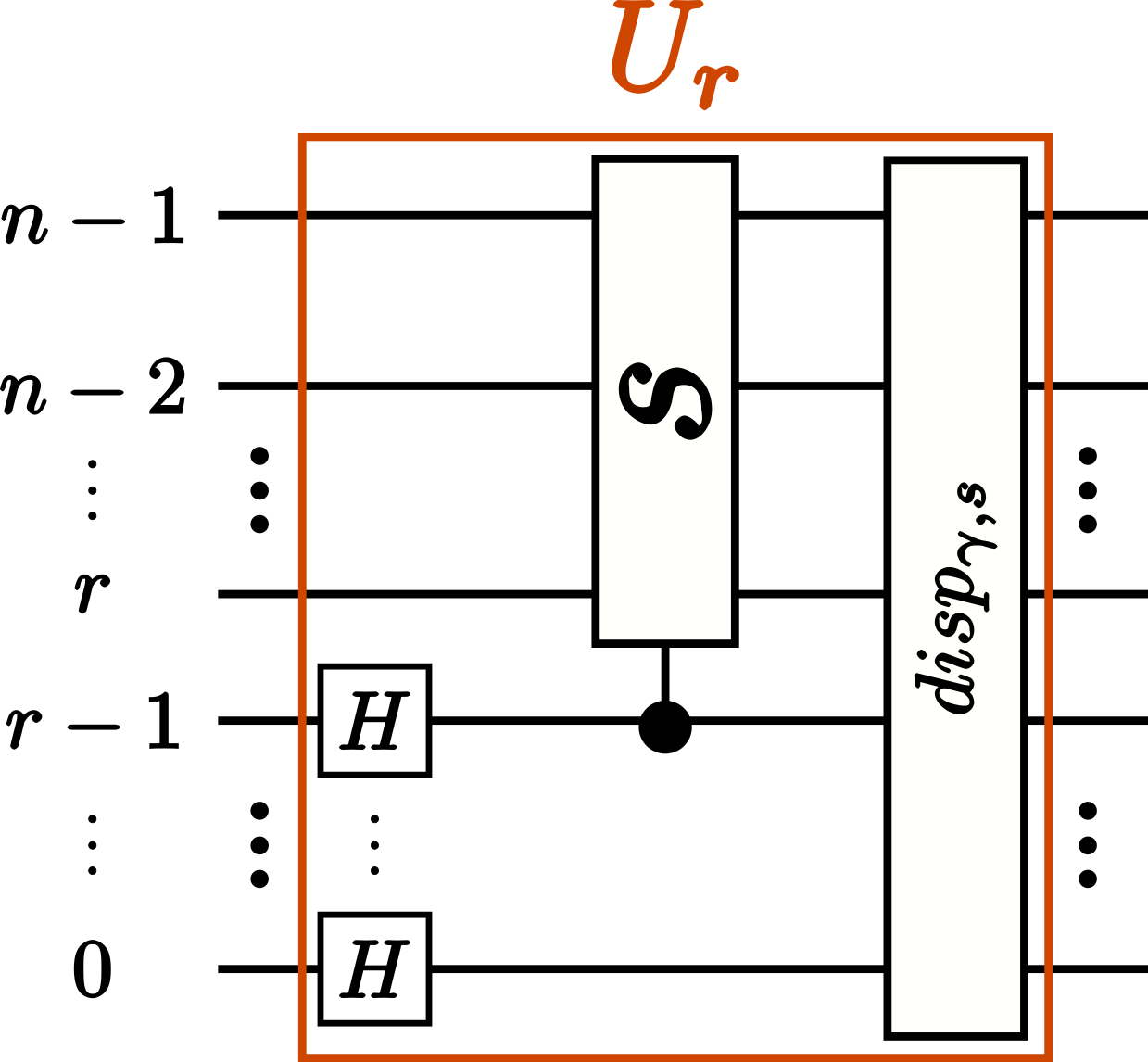}
    \captionof{figure}{Range splitter operator}
    \label{fig:range-splitter-operator}
\end{Figure}
\end{widetext}
This operator superposes a total of $2^{r}$ states in two eventually non-consecutive ranges. In fact, they will be located at a distance of $s\cdot 2^{r}$ states, where $s\in\mathbf{N}$.

This operator first superposes some qubits in order to generate the two ranges which relative location is the following one:

\begin{equation}\label{q:range-r-t}
    [0,2^{r-1}-1]\cup[(s+\frac{1}{2})\cdot 2^{r}, (s+\frac{1}{2})\cdot 2^{r} + 2^{r-1}-1]
\end{equation}
Fig. \ref{fig:example-range-superposition} shows these pairs of ranges (after the corresponding displacement which will be described below) in their absolute (proper) location around $\ket{\gamma}$. Please notice that although $\ket{\gamma}$ corresponds to $10^{th}$ state, for the sake of keeping pure disjoint ranges the exact median over which we have defined and built our ranges is an state between $10^{th}$ and $11^{th}$ ones.
\begin{widetext}
\begin{Figure}
    \centering
    \includegraphics[width=\linewidth]{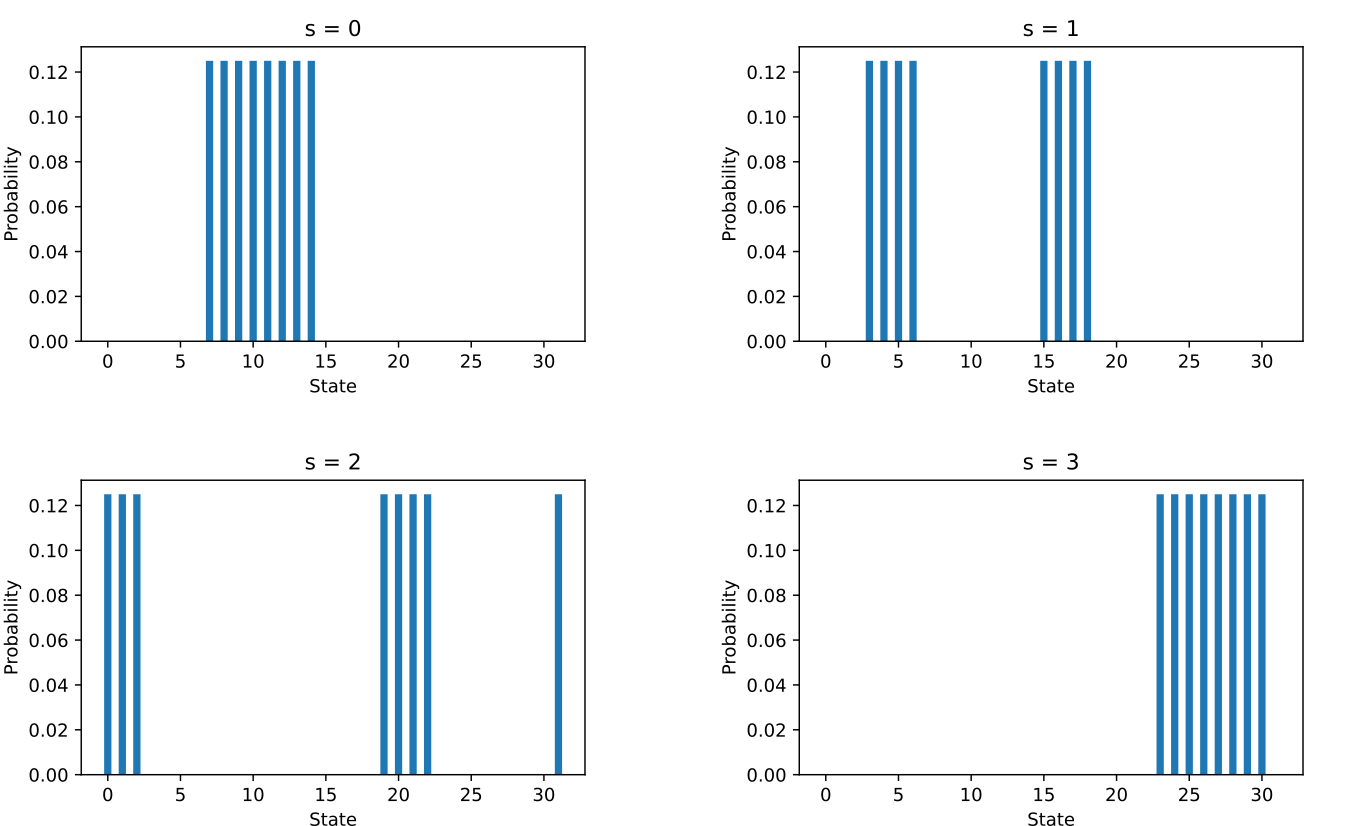}
    \captionof{figure}{Example of superposition with $\gamma=10$, $r=3$, $n=5$ and $s=0,1,2,3$}
    \label{fig:example-range-superposition}
\end{Figure}
\end{widetext}
In contrast with the most of the iterative quantum searching methods, our proposal fully avoids  repetitions, i.e., it ranges over disjoint domains. Let's see on detail how does this superposition generates more and more distant ranges of neighbours, is achieved:

\textit{Hadamard} gates in fig. \ref{fig:range-splitter-operator} create a superposition of the first (from least to most significant) $2^{r}$ states:
\begin{equation}
    \sum_{x_0,\ldots, x_{r-1} \in\{0,1\}}\frac{1}{\sqrt{2^{r}}}\cdot\underbrace{\ket{00\ldots 0x_{r-1}\ldots x_{0}}}_{n}
\end{equation}

First the biggest $2^{r-1}$ states, i.e., $(\ket{00\ldots 0\textbf{1}x_{r-2}\ldots x_{0}})$ are shifted by an amount of $s\cdot 2^{r}$.
Performing this addition makes use of quantum parallelism in the following way. The states to which we need to perform the addition (most significant states) have in common that their qubit in $r^{th}$ position, $(x_{r-1})$, is set to $\ket{1}$. This qubit will act as control qubit for adding the value $s\cdot 2^r$ only to the most significant qubits.

These additions make use of a set of \textit{full adders}\footnote{Logical circuits that add two bits and an input carry in order to produce a sum and an output carry} to perform the operation.

Taking into account that: \begin{itemize}
    \item Full adders are performed by means of \textbf{XOR} $(\oplus)$ operator (which neutral element is $0$).
    \item We are adding $s\cdot 2^r$ (which is a multiple of $2^r$ and therefore its $r-1$ least significant qubits are set to $0$)
    \item Full adders are applied to the $n-r$ most significant qubits which are set to $0$ $(\ket{00\ldots 0\textbf{1}x_{r-2}\ldots x_{0}})$
    \end{itemize}

Value $s\cdot 2^{r}$ can be added to the values $0$ just 
setting the values $1$ with controlled $X$ gates on the $n-r$ most significant qubits. These values $1$ correspond to those values $1$ appearing in the binary representation of $s\cdot 2^{r}$. Previous computations are equivalent to simply set the value of $s$ (binary encoded) on the $n-r$ most significant qubits (second column in fig \ref{fig:range-splitter-operator}).

Only displacements remain in order to finish $U_r$ performing description. 
To do this, at iteration $s$, we need to perform a displacement (which depends on the QA output state $\ket{\gamma}$) of $disp_{\gamma,s}=\gamma - \left(s+1\right)\cdot 2^{r-1}+1$ This is achieved by means of performing $(+1)$ quantum gates that operates $+1\ mod\ 2^{n}$ on an $n$-qubit register. These increments are carried out as described in \cite{incremental-gate}.

On practice, $disp_{\gamma,s}$ is decomposed into as many increments as values $1$ appear within the binary representation of the displacement. Each of these qubits $(x_{i})$ will apply the increment quantum gate shown in fig.\ref{fig:incremental-gate} in the way which is described in fig.\ref{fig:displacement-gate}. As an example figs. \ref{fig:example-displacement-circuit} and \ref{fig:example-displacement-plot} show on detail the case where the initial state is 13 $\ket{01101}$ and a displacement of 5 is required $(x_{4}x_{3}x_{2}x_{1}x_{0}=00101)$. The number of increments required for this equals the number of variables set to 1, which is 2 (coming from $x_{2}=x_{0}=1$). Regarding how these increments must be performed let's say that applying the increment corresponding to $x_{2}=1$ results in an increment for the three most significant qubits, whereas the increment corresponding to $x_{0}=1$ results in an increment for the five most significant qubits, i.e., for all the qubits under consideration.


\begin{widetext}
\begin{Figure}
    \centering
    \includegraphics[width=0.75\linewidth]{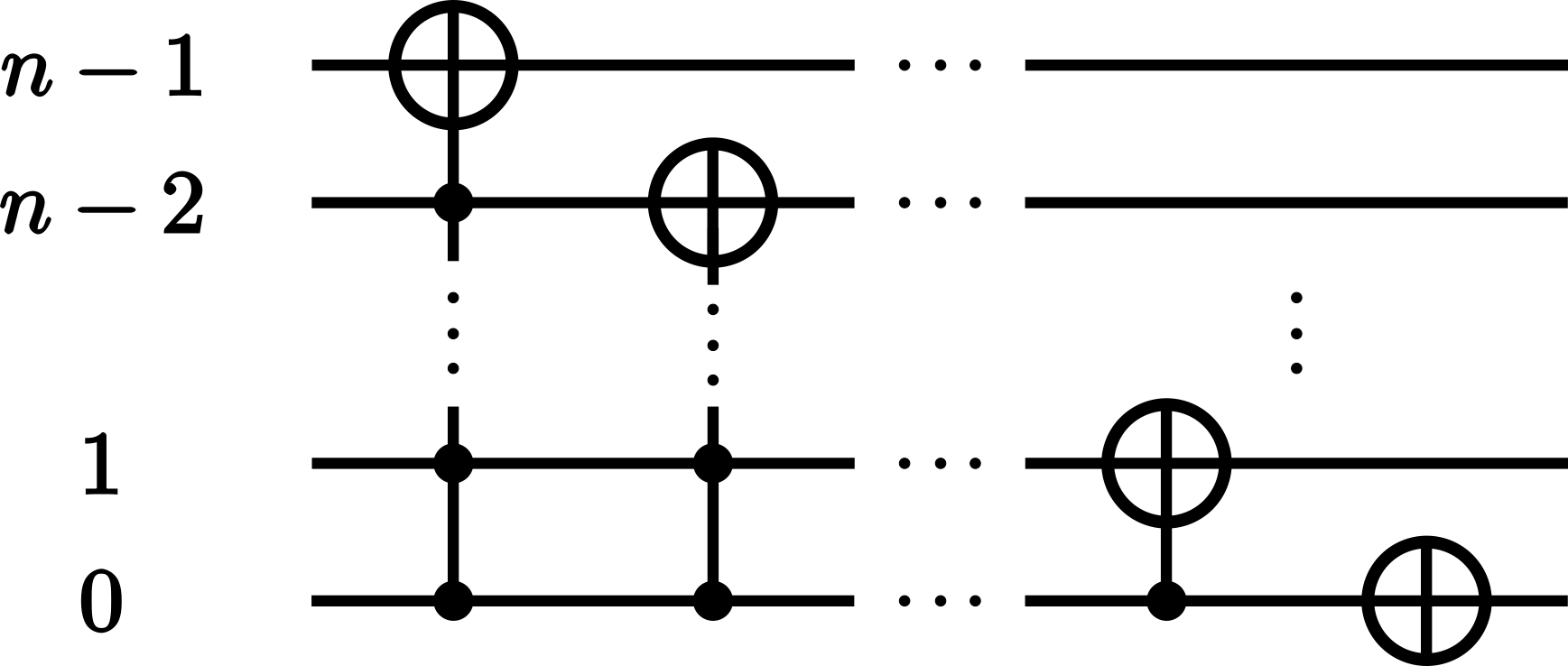}
    \captionof{figure}{Incremental gate circuit}
    \label{fig:incremental-gate}
\end{Figure}
\begin{Figure}
    \centering
    \includegraphics[width=\linewidth]{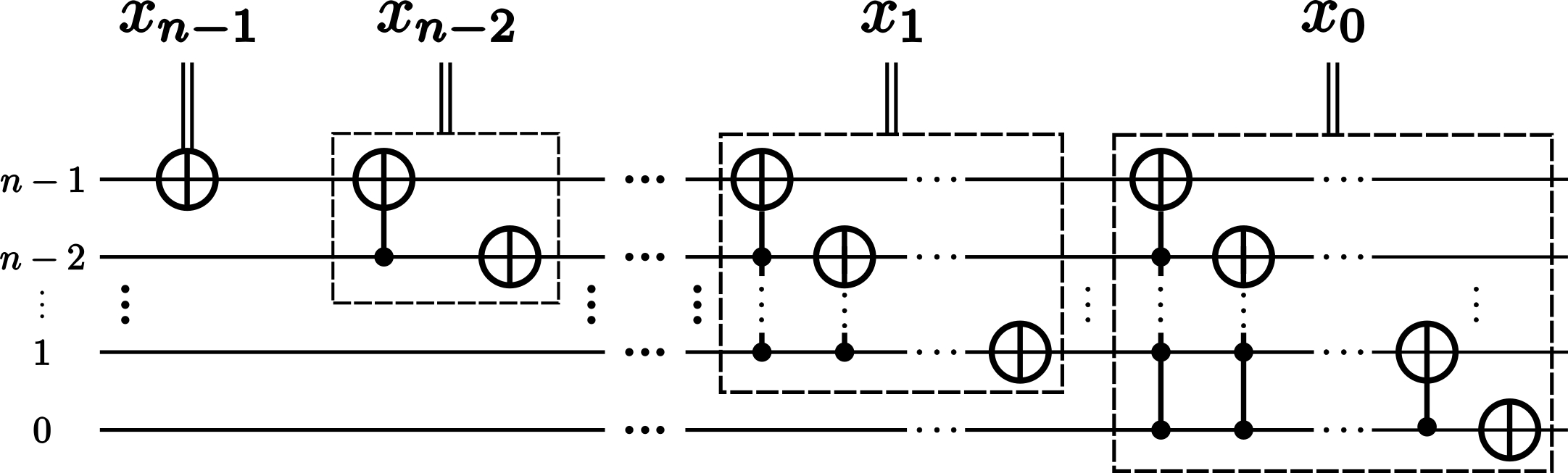}
    \captionof{figure}{Displacement gate circuit made of Incremental gates ones}
    \label{fig:displacement-gate}
\end{Figure}

\begin{Figure}
    \centering
    \includegraphics[width=0.8\linewidth]{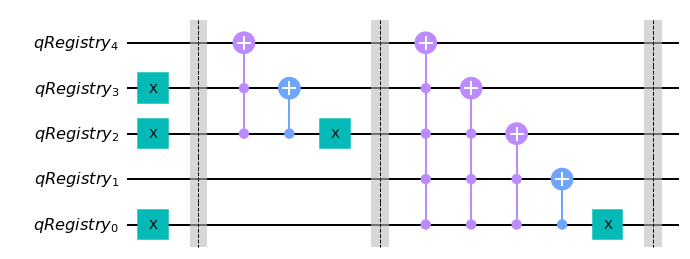}
        \captionof{figure}{Example of a \textit{Displacement quantum gate circuit} over Qiskit.}
    \label{fig:example-displacement-circuit}
\end{Figure}
\begin{Figure}
\centering
    \includegraphics[width=0.7\linewidth]{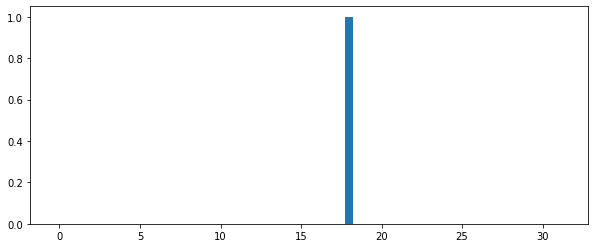}
    \captionof{figure}{The final probability distribution of the example over Qiskit.}
    \label{fig:example-displacement-plot}
\end{Figure}
\end{widetext}
$U_r$ operator so defined is used by the inversion about the mean $\left(IAM_{r}\right)$ operator as follows:
\begin{equation}
    IAM_{r}=2\cdot U_{r}\ket{0}\bra{0}U_{r}^{\dagger}-I_{n}
\end{equation}
Fig. \ref{fig:ism-cyclic} depicts, on detail, its corresponding quantum circuit, once again the suitable -1 global phase has been added.
\begin{widetext}
\begin{Figure}
    \centering    \includegraphics[width=0.82\linewidth]{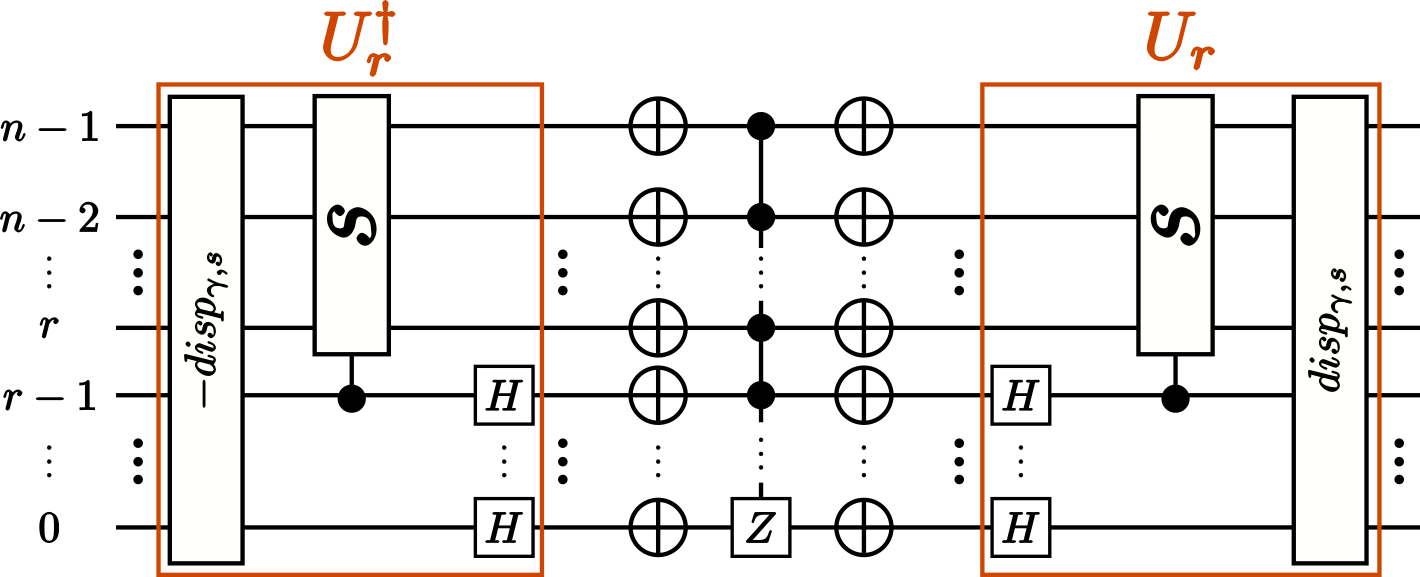}
    \captionof{figure}{Inversion about the mean (Cyclical distance)}
    \label{fig:ism-cyclic}
\end{Figure}
\end{widetext}
The algorithm we propose consists on iterating $t_{r}$ times the operator $G_{r}$, that is defined as:
\begin{equation}
    G_{r}=\left(2\cdot U_{r}\ket{0}\bra{0}U_{r}^{\dagger}-I_{n}\right)\cdot U_{f}
\end{equation}

The number of iterations $t_{r}$ needed to reach a probability of measuring the solution state close to 1 can be computed as:
\begin{equation}
    t_{r}=\frac{\pi}{4\left(sin^{-1}\left(\sqrt{\frac{|S|}{N}}\right)\right)}-\frac{1}{2}
\end{equation}
Where $N$ is the total number of states in superposition, $|S| = 1$ as previously stated is the total number of solutions. Notice that the total number of states value comes from $N=2^{r}$. Therefore we get:
\begin{equation}
    t_{r}=\frac{\pi}{4\left(sin^{-1}\left(\sqrt{\frac{1}{2^{r}}}\right)\right)}-\frac{1}{2}
\end{equation}
from which the number of iterations required is $Round(t_r)$, i.e. $\lfloor t_r \rceil$.

Again, $\lfloor t_r \rceil$ are the values to be added in order to get the key element to be compared.

\section{Testing scenario}\label{subsec:annealing-circuit-testing-the-proposal}
For the sake of giving the most details for reproducibility this section is divided into:
\begin{itemize}
    \item Test cases
    \item QUBO models \& Quantum Annealing
    \item Quantum Hamming and Cyclical searches
\end{itemize}
\subsection{Test cases}
Some preliminary definitions are required:
\begin{itemize}
    \item The density of a given k-SAT formula $S_{f}=\bigwedge_{i=1}^{i=m}C_{i}$ is denoted as $d(S_{f})$ and defined as the ratio between its number of clauses (m) and its number of Boolean variables (n): $\displaystyle d(S_{f})=\frac{m}{n}$
\end{itemize}

On the related literature it is assumed that $\forall k>1, k \in \mathbb{N}$ there exists a threshold value for the density of k-SAT formulas $d_k$ such that:

\begin{itemize}
    \item The farther \(d(S_{f})\) is from \(d_{k}\), the fewer calls to the DPLL algorithm \cite{Davis1962ATheorem-proving}) are required to solve the k-SAT problem.
    \item If \(d(S_{f}) > \ d_{k}\) the formula would be unsatisfiable with high probability but just the opposite occurs when \(d(S_{f}) < d_{k}\), i.e., high $d(S_{f})$ is usually associated with an unsatisfiable formula.
    \item Mitchell, Selman and Levesque estimated in 1991, see \cite{sat-threshold}, that $d_3 \sim 4.55$ for about 20 variables and $d_3 \sim 4.3$ for larger number of variables.
\end{itemize}


Regarding our scenario for the tests:

\begin{enumerate}
    \item The process of generating test cases has been supported by \textbf{CNFgen}\cite{cnfgen}. A test case for us is a pair (density, number of variables). In particular, 10 3-SAT problems have been generated for each test case. 
    
    Supported by those results by Mitchell et al., \cite{sat-threshold}, we have considered 4, 4.3 and 4.55 as most challenging density values.
    
    The number of variables ranges from 7 to 22 (for less than 7 variables annealing algorithms reach the 3-SAT solution easily by themselves whereas the upper threshold of 22 variables has been chosen because together with the high computational cost, we have found neither evidence nor hints that make us suspecting that the advantage we would get beyond 22 could be smaller). Therefore, it computes $3 \cdot 16$ different test cases which computes in total 480 3-SAT problems.
    \item In addition to these 480 3-SAT problems, 5 random seeds has been also involved within the setup for the sake of maximum representativity, thus $5 \cdot 10 \cdot 3 \cdot 16 = 2400$ different executions over simulated annealing algorithm have been performed.
\end{enumerate}

The number of iterations required by the algorithm to reach a probability of measuring a solution state close to 1, has been chosen as the parameter of study in order to compare the  computational efficiency of the proposed algorithm.

The scenarios we had in mind at the beginning were the first 3 in tab. \ref{tab:scenarios} but due to the low quality of the quantum annealing solutions we have added the last 2 which cover simulated annealing instead of quantum annealing.

\begin{widetext}
\begin{table}[h]
\begin{tabular}{|l|l|l|}
\hline
 Scenario & Annealing algorithm & Circuit quantum search algorithm  \\
 \hline
1 & Not applicable & Grover's algorithm\\
2 & Quantum annealing & Quantum Hamming search  \\
3 & Quantum annealing & Quantum Cyclical search \\
4 & Simulated annealing & Quantum Hamming search \\
5 & Simulated annealing & Quantum Cyclical search \\
\hline
\end{tabular}
\caption{Scenarios meaning}
\label{tab:scenarios}
\end{table}
\end{widetext}

\subsection{QUBO models \& Quantum Annealing}

The QUBO models and the results we have generated and analysed are available in this repository \cite{Paulet2023}.

This annealing algorithms part just seek to save temporal complexity on the pure quantum computing searching process. 
To do this, we have calculated the average time it took to execute both annealing algorithms (quantum and classical) on the 10 examples of each test case.

For both quantum annealing and simulated annealing, the temporal complexity is linear as we can see in detail in the Appendices \ref{anexo:dwave-linear} and \ref{anexo:simulated-linear}, respectively.

For performing these experiments, we have made use of the software development kit (SDK) provided by DWave (Ocean). Specifically, for the simulated annealing algorithm, we have used the \textit{dwave-neal} library and its sampler \textit{SimulatedAnnealingSampler}. For the quantum annealing execution, we have used the cloud service \textit{Leap} from DWave along with its SDK, and the \textit{dwave-system} library, which contains the \textit{DWaveSampler} for running on the Quantum Processing Units (QPUs).


To calculate the number of iterations that we would perform in each of the 2 versions of quantum search without simulating the circuits (since we cannot simulate them for more than 6 variables), we will calculate the Hamming distance $(k_{f})$ and cyclical distance $(d_{f})$ between the solution provided by the annealing algorithms and the real solution to the problem in question. Thus, once this distance is known we will be on condition to calculate  the number of iterations that we will perform in each case to obtain the real solution starting from the solution provided by the annealing stage.


\subsection{Applying Quantum Hamming search}
As previously stated the number of Grover iterations to be performed $t_{\alpha}$ depends on $\displaystyle \frac{n}{k}$ and on $k_f$. For estimating $k_{f}$ we will iteratively execute the algorithm for the values of $k$ in $\{1,\ldots,k_{i},\ldots,k_f\}$ until we get a probability of measuring the solution state greater or equal than 90\%.

The probability of measuring the solution state within the first $k_i$ attempts $(k_i\leq k_f)$ where the distance to actual solution is $k_f$, is referred as $P_{sol}\left(k_{i},k_{f},n\right)$ and computed in this way:
\begin{widetext} 
\begin{equation}
    P_{sol}\left(k_{i},k_{f},n\right)=1-\prod_{k=1}^{k_{i}}\left(\overbrace{1-\underbrace{\left(\sin\left(\left(2\cdot \lfloor t_{\alpha}\rceil+1\right)\cdot\sin^{-1}\left(\sqrt{\left(1-\frac{k}{n}\right)^{n-k_{f}}}\cdot\sqrt{\left(\frac{k}{n}\right)^{k_{f}}}\right)\right)\right)^{2}}_{solution \ probability}}^{non \ solution \ probability}\right)
\end{equation}
\end{widetext}
Where $t_{\alpha}$ comes from equation \ref{eq:iterations-hamming} and $\lfloor\chi\rceil$ stands for $Round(\chi)$ (round approximation).

Due to the fact that the probability of measuring one state or another does not depend on previous runs of the algorithm, the probability after $k_{i}$ executions equals 1 minus the product of measuring a non-solution in each of the previous independent events (each algorithm's run). Therefore, the total number of iterations needed in a real case according to this Hamming search will be:
\begin{widetext}
\begin{equation}    t_{H_{total}}=\sum_{k=1}^{k_{i}\leq k_{f}}\left\lfloor\frac{\pi}{4\cdot \sin^{-1}\left(\sqrt{\left(1-\frac{k}{n}\right)^{n-k_{f}}}\cdot\sqrt{\left(\frac{k}{n}\right)^{k_{f}}}\right)}-\frac{1}{2}\right\rceil
\end{equation}
\end{widetext}
Where $k_{i}=\mu z.P_{sol}\left(z,k_{f},n\right)\geq 0.9$ where $z \in \mathbb{Z}$.
\subsection{Applying Quantum Cyclical search} \label{QCS}
For the quantum search algorithm based on a sort of cyclical distance, we first need to determine the number of qubits to be superposed, $r$ out of the total number of qubits in the system $n$ (where $N=2^n$ is the total number of states under consideration). We would like to set $r$ to the minimum value which generates a range that can reach the solution state from the state provided by QA, $\ket{\gamma}$.

The cyclical distance we are considering is defined as follows:
\begin{equation}\label{eq:cyclic-distance}
    d_c(\ket{\gamma},\ket{\tau})=min\left\{\left|\gamma_{2}-\tau_{2}\right|,2^{n}-\left|\gamma_{2}-\tau_{2}\right|\right\}
\end{equation}

For example, 
\begin{equation}
\begin{split}
   & d_c(\ket{0010},\ket{1101}) =  \\ & = min\left\{\left|(0010)_{2}-(1101)_{2}\right|,\right. \\
   & \ \ \ \ \ \ \ \left. 2^{4}-\left|(0010)_{2}-(1101)_{2}\right|\right\} =\\ 
   & = min\left\{\left|2-13\right|,16-\left|2-13\right|\right\} = 5 
\end{split}
\end{equation}
which is graphically depicted in fig.\ref{fig:cycle-distance}.  

\begin{Figure}
    \centering
    \includegraphics[width=0.5\linewidth]{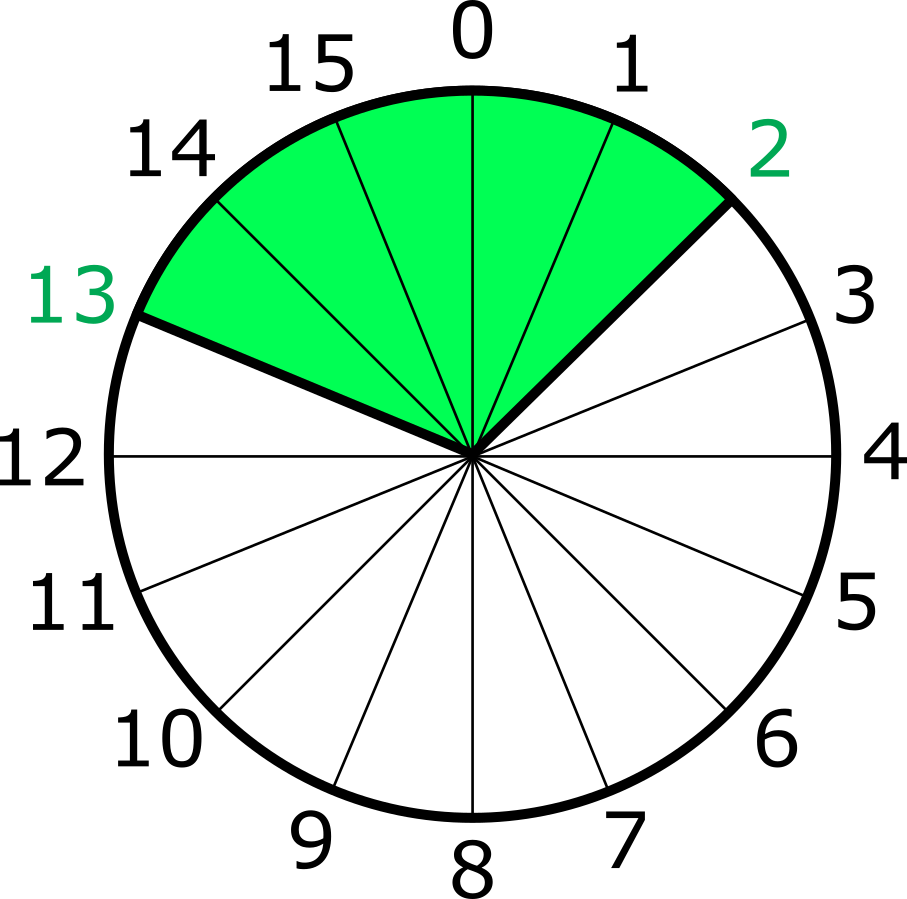}
    \captionof{figure}{Cyclical distance between  $\ket{0010}$ and $\ket{1101}$}
    \label{fig:cycle-distance}
\end{Figure}

As, obviously, the location of the solution state is not known until the algorithm has finished but we need to estimate the total number of iterations indeed, we are forced to find the solution of the SAT problem by brute force, afterwards computing the cyclical distance between it and the output of QA, $d_f$, in order to keep on with this comparative study. 

As stated we are required to know the total number of iterations $t_{C_{total}}$ in each test case, for which we start  setting $r=n-1$ (which covers half of the whole domain per execution). Starting with $r=n-k$ (covering a $1/2^k$ of the whole domain per execution) conforms other possible options. Our decision is just supported by empirical issues. 




Therefore, $d_f$ must be divided by the number of states covered in any of the ranges previously defined. Thus, from state $\ket{\gamma}$ 
we cover $s\cdot 2^{r-1}-1$ states to the left and $s\cdot 2^{r-1}$ states to the right (because of symmetry issues already pointed out in previous section \ref{subsec:quantum-cyclic-search}). Therefore, the number of algorithm executions $s_{f}$ to be performed (corresponding to former parameter $s$) is:
\begin{equation}
    s_{f}=\left\{
    \begin{matrix}
    \left\lceil\frac{d\left(\gamma,\tau\right)+1}{2^{r-1}}\right\rceil & if\ \gamma\geq \tau\\
    \left\lceil\frac{d\left(\gamma,\tau\right)}{2^{r-1}}\right\rceil & if\ \gamma<\tau
    \end{matrix}\right.
\end{equation}

Consequently, the total number of iterations will be:

\begin{equation}
    t_{C_{total}}=\sum_{i=0}^{s_{f}}\left\lfloor \frac{\pi}{4\left(sin^{-1}\left(\sqrt{\frac{1}{2^{r}}}\right)\right)}-\frac{1}{2}\ \right\rceil
\end{equation}

This parameter $t_{C_{total}}$ together with $t_{H_{total}}$ are the final parameters that we understand best describe the computational cost of both Quantum Search proposals, and therefore the right ones to be compared.

\section{Summary of Results}\label{subsec:annealing-circuit-results}

To begin with, we want to mention that as it is well known that the hard achievability to actual quantum computers in addition to the strong dependency on the topology of the computer, on the circuit optimization, as well as on the characteristics of the quantum gates makes almost impossible to empirically mease times of execution. Instead, the number of iterations within the Grover's alike algorithm are considered as proper size of the problem for comparisons.

Besides, DWave's quantum annealers execution time depends on two main parameters: the $\textit{anneal\_schedule}$ (or \textit{annealing\_time}), and the number of executions (\textit{num\_reads}). Similarly to quantum circuit model the variability associated to the offer of hardware makes almost impossible to establish criteria in order to accurately compare efficiency of these two different quantum computing models.

In front of the previous scenario we have focused on the fact that quantum searches grows exponentially meanwhile quantum annealing grows linearly, therefore we consider as main parameter to be considered for comparisons just the usual one for this task within circuit model quantum search, i.e. the number of Grover's iterations. 


Some preliminary considerations regarding quantum annealing and simulated annealing must be stated.
We started evaluating the quality/accuracy of these annealing processes according to both Hamming and Cyclical metrics. In order to do this, we compare the distances from the solution state offered by QA $\ket{\gamma}$ and the actual solution state of the former 3-SAT problem $\ket{\tau}$. 


\begin{widetext}
\begin{Figure}
    \centering
    \includegraphics[width=0.8\linewidth]{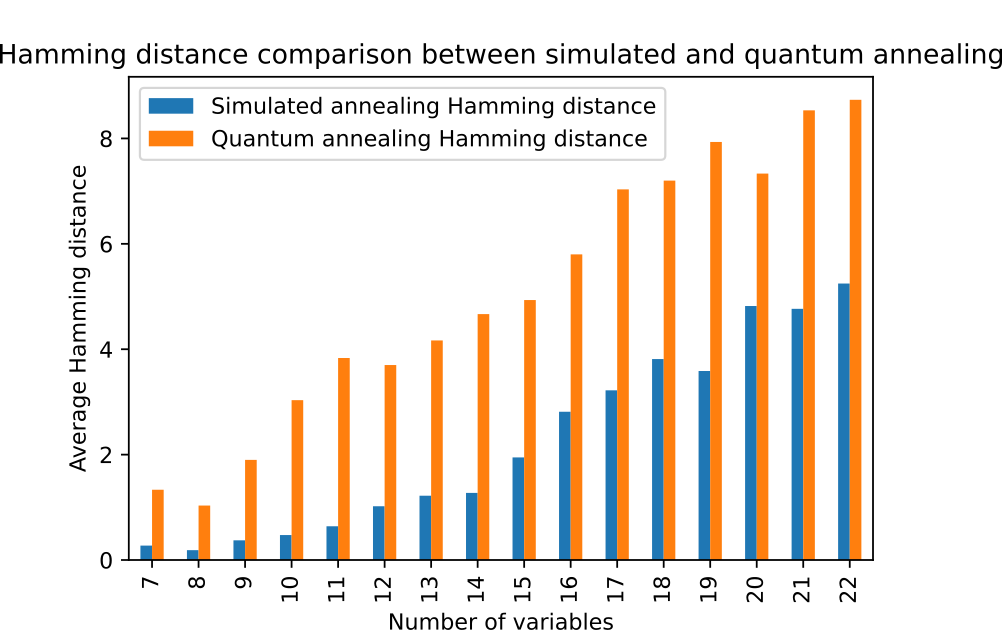}
    \captionof{figure}{ Simulated annealing vs. Quantum annealing Hamming accuracy}
    \label{fig:hamming-distance-comparison}
\end{Figure}
\end{widetext}
As fig.\ref{fig:hamming-distance-comparison} shows Simulated Annealing is far more accurate than Quantum Annealing according to Hamming distance. This is quite remarkable as it strongly affects the subsequent Quantum Search.

Fig.\ref{fig:decimal-distance-comparison} compares that accuracy with respect to Cyclical distance. Let us notice that vertical axe follows a logarithmic scale.
\begin{widetext}
\begin{Figure}
    \centering
    \includegraphics[width=0.8\linewidth]{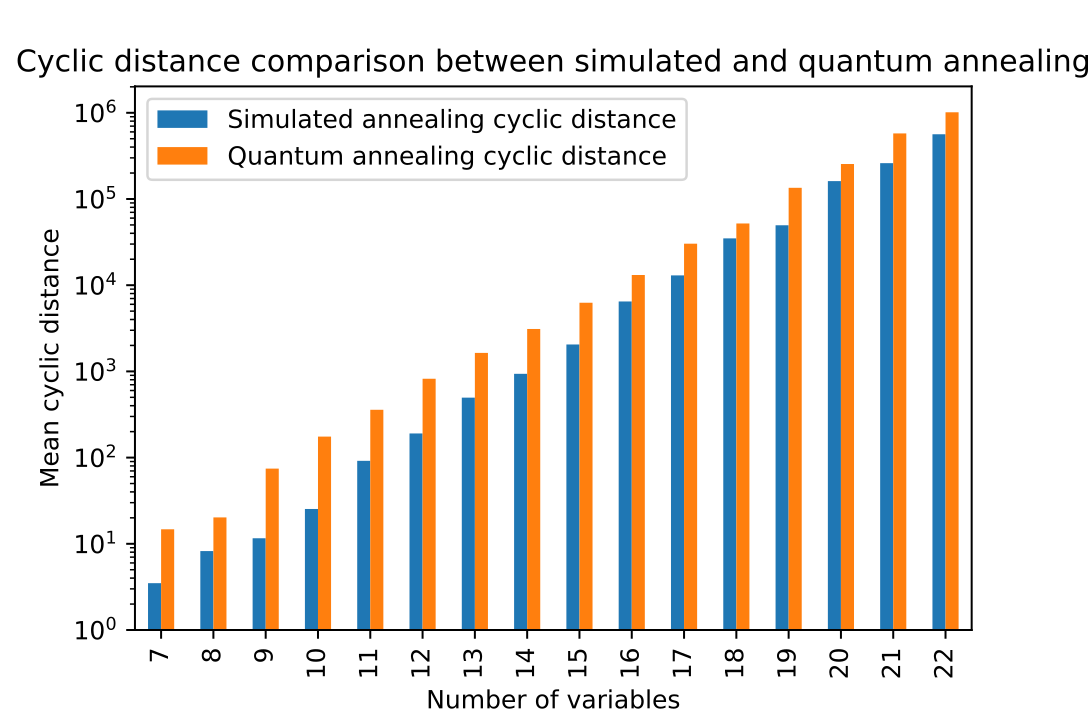}
    \captionof{figure}{Simulated annealing vs. Quantum annealing Cyclical accuracy}
    \label{fig:decimal-distance-comparison}
\end{Figure}
\end{widetext}
In the same way, Simulated Annealing clearly overcomes Quantum Annealing accuracy according to Cyclical distance.

More detailed figures of the experiments performed for this sake can be found in the table within appendix \ref{anexo:results-quantum-simulado-distance}.

Previous results leaded us to focus on scenarios 1, 4, and 5 of tab.\ref{tab:scenarios} until any further Quantum Annealer improves the accuracy of the Simulated Annealing at disposal.

\begin{widetext}
\begin{Figure}
    \centering
    \includegraphics[width=0.8\linewidth]{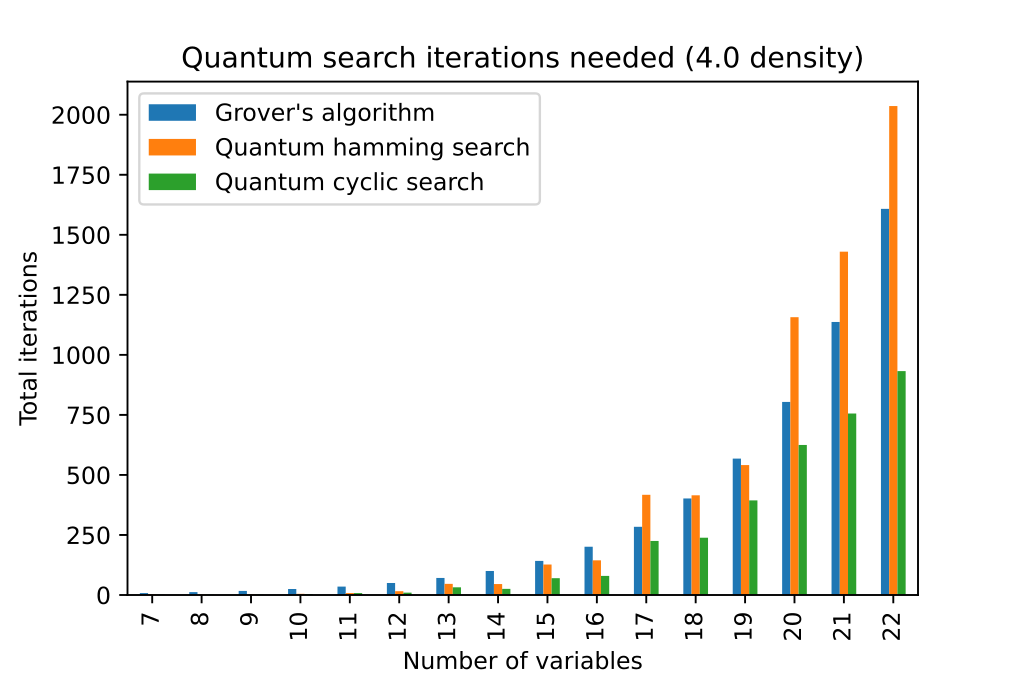}
    \captionof{figure}{Total number of iterations required in scenarios 1, 4 and 5 over a 3-SAT formula with density 4.0}
    \label{fig:iterations-comparison-4}
\end{Figure}
\end{widetext}

\begin{widetext}
\begin{Figure}
    \centering
    \includegraphics[width=0.8\linewidth]{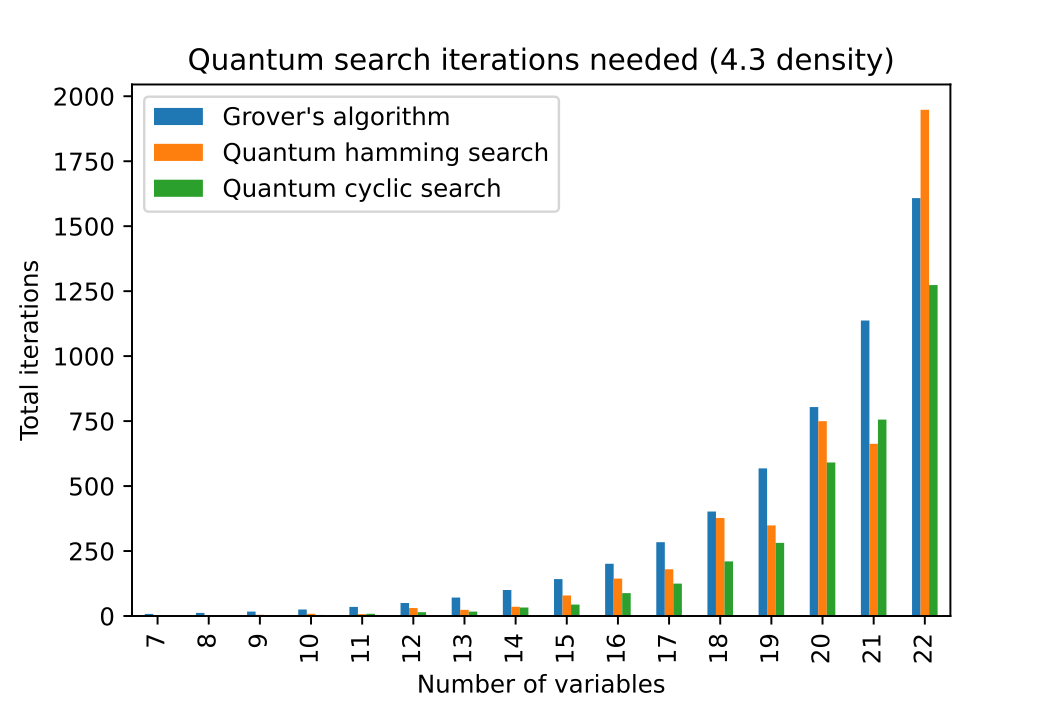}
    \captionof{figure}{Total number of iterations required in scenarios 1, 4 and 5 over a 3-SAT formula with density 4.3}
    \label{fig:iterations-comparison-4.3}
\end{Figure}

\begin{Figure}
    \centering
    \includegraphics[width=0.8\linewidth]{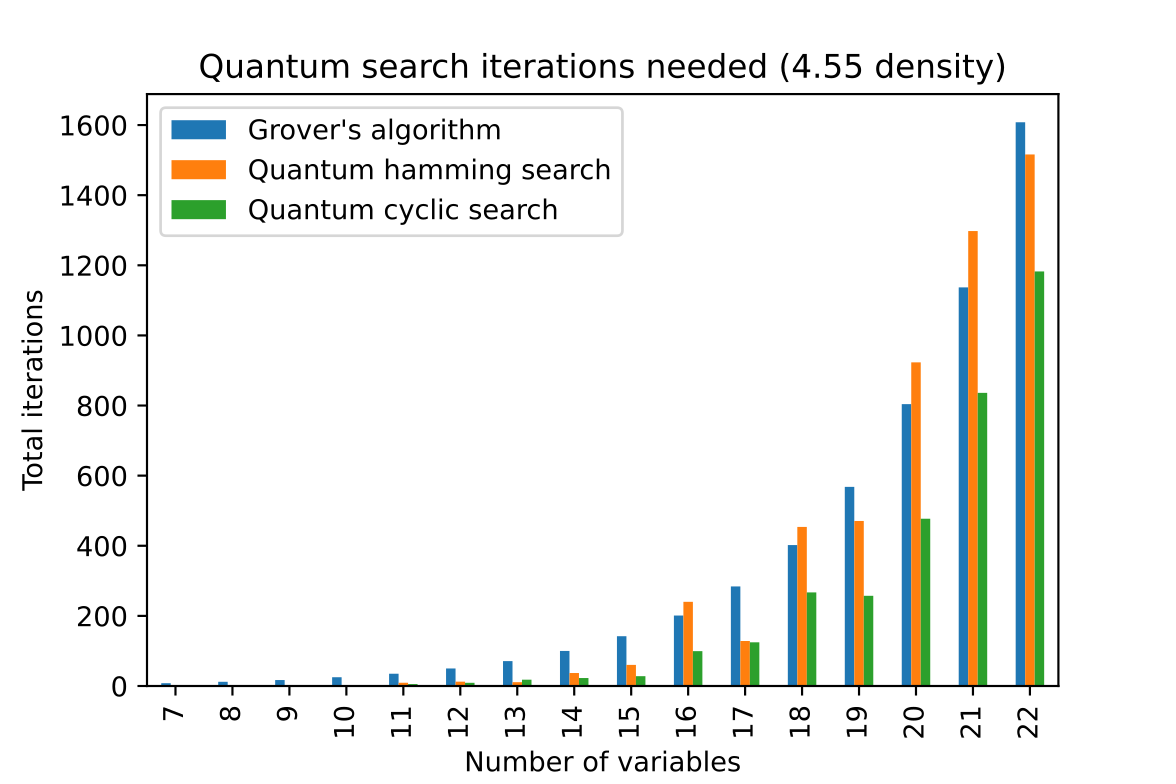}
    \captionof{figure}{Total number of iterations required in scenarios 1, 4 and 5 over a 3-SAT formula with density 4.55}
    \label{fig:iterations-comparison-4.55}
\end{Figure}
\end{widetext}

Graphs depicted on figs. \ref{fig:iterations-comparison-4}, \ref{fig:iterations-comparison-4.3} and \ref{fig:iterations-comparison-4.55} summarize, stratified by density of the SAT formulas under consideration, the results obtained in our test cases for comparing the two proposed metrics. We can conclude from them on one hand that Hamming search proposal does not improve in general the former Grover quantum Search, on the other hand Cyclical search proposal clearly improves Grover's algorithm performance though.

We think that the main reason for this relies on the disjoint domains search which is just achievable by cyclical search.

Appendix \ref{sec:anexo-model-circuit-results} includes tabs. \ref{tab:total-iterations-7-14} and \ref{tab:total-iterations-15-22} which contain the raw numbers obtained in the comparative tests that we have performed.

Such improvement reaches in the very worst case, i.e. density around 4.3, a saving of one fifth of the number of iterations with respect to the former Grover algorithm.

\section{Conclusions and future work}

The main contributions of the paper are:
\begin{itemize}
    \item We have provided a hybrid quantum computing proposal to deal with 3-SAT problem. 
    \item We have provided and compared a couple of metrics which generate a couple of circuit model quantum searching processes the Grover's one alike, i.e. Quantum Hamming and Cyclical Searches.
    \item We have concluded that it clearly deserves to first approximate 3-SAT problem solving by quantum annealing sampling and then apply quantum cyclical search than just applying Grover's algorithm from scratch.
\end{itemize}

For future work we think that it could be interesting:
\begin{itemize}
    \item Studying and developing heuristic approaches for selecting the appropriate values of the distance parameters ($k$ and $r$) in Quantum Hamming search and Quantum Cyclical search algorithms, respectively.
    \item Investigating the Reverse Annealing Technique as it empowers users to define the problem they intend to solve and to provide an anticipated solution, thereby narrowing down the search space during the computation process. 
    \item Exploring Anneal Offsets because in some problem scenarios, it can result in an advantage for specific qubits to undergo annealing slightly before or after others. 
\end{itemize}

\bibliographystyle{quantum}
\bibliography{quantum-template}

\onecolumn
\appendix

\section{DWave's quantum annealing belongs to O(n)}\label{anexo:dwave-linear}

As shown in~\cite{qa-sa-equivalence}, simulated annealing and quantum annealing are conceptually equivalent. 

\begin{figure}[h]
\centering
\includegraphics[width=0.8\textwidth]{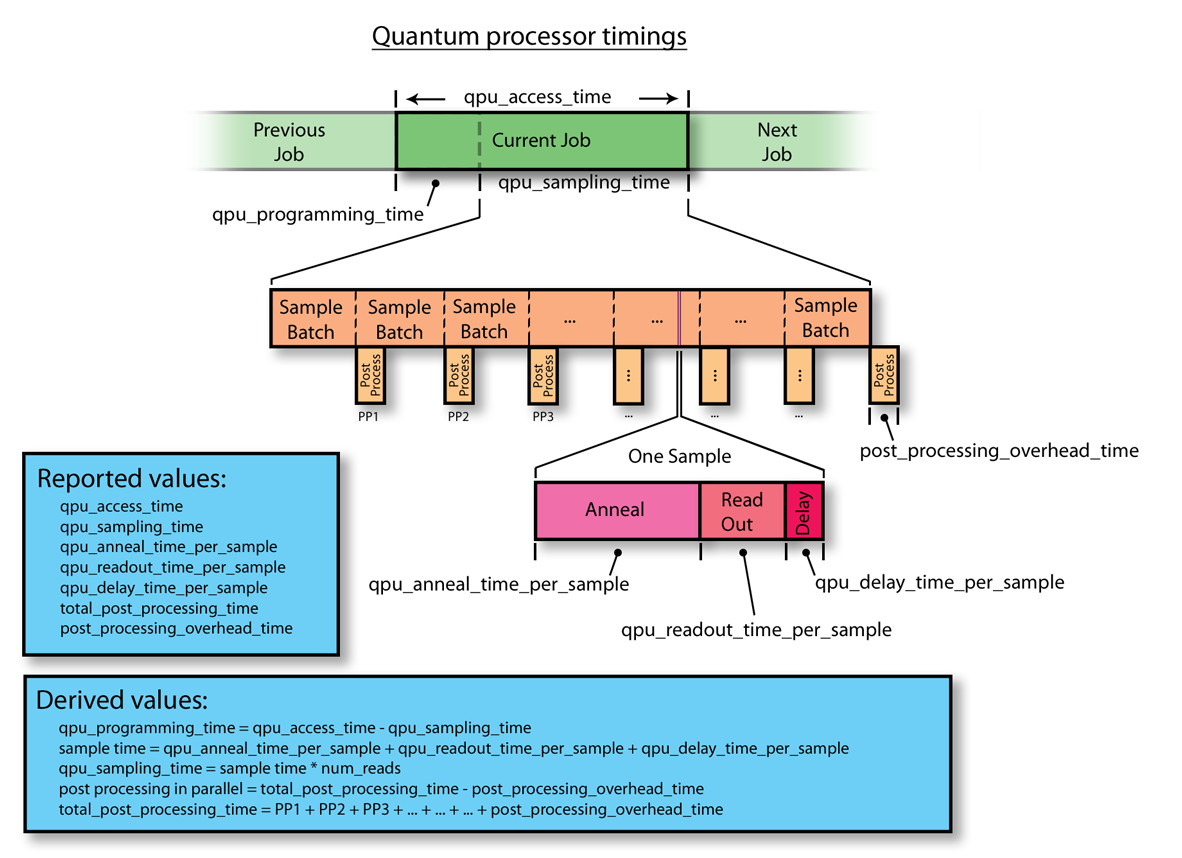}
\caption{D-Wave QPU timing structure. Image from https://docs.dwavesys.com/}
\label{fig:dwave-timming-structure}
\end{figure}
\begin{figure}[h]
\centering
\includegraphics[width=0.4\textwidth]{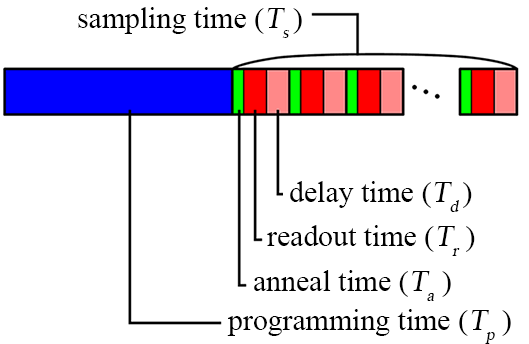}
\caption{Detail of QPU access time. Image from https://docs.dwavesys.com/}
\label{fig:dwave-qpu-access-time}
\end{figure}

Each task performed on DWave quantum annealers takes a total time of \textit{qpu\_access\_time}. This time is divided into \textit{qpu\_programming\_time} and \textit{qpu\_sampling\_time} as we can see in fig. \ref{fig:dwave-timming-structure}. The \textit{qpu\_programming\_time} is the time performed once at initializing stage of the QPU (regardless the number of measured qubits or the number of executions or any other issue), and the \textit{qpu\_sampling\_time} is the time for multiple sampling times within the current execution on the QPU. Fig. \ref{fig:dwave-qpu-access-time} shows a more detailed information of it.

For that, the resulting \textit{qpu\_access\_time} is:

\begin{equation}
    T=T_{p}+\Delta+T_{s}
\end{equation}
Where $T_{p}$ is  \textit{qpu\_programming\_time} and $T_{s}$ is \textit{qpu\_sampling\_time}. $\Delta$ stands for an initialization time spent in low-level operations.

We estimated that the most accurate hardware option within D-Wave offer is \textit{DW\_2000Q\_6}.  
In fact, as shown in \cite{2000q-better-results-sparse-problems} \textit{D-Wave 2000Q} systems overcome \textit{D-Wave Advantage} at accuracy when dealing with sparsely connected problems as for example that which is depicted in fig.  \ref{fig:7_vars_sparse_problem} which corresponds to one of our examples with 7 variables. Let's notice that although the maximum number of variables in our work reaches 22 the density prevent them from being non-sparse.
\begin{Figure}
    \centering
    \includegraphics[width=0.8\linewidth]{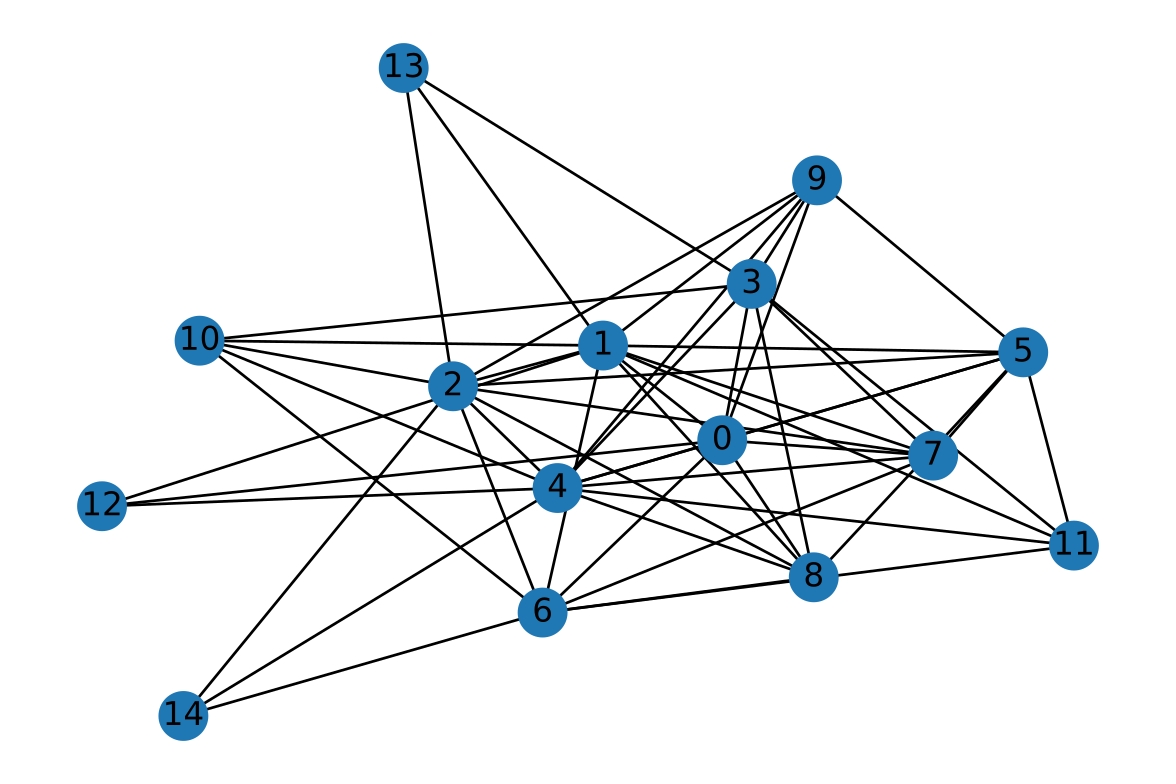}
    \captionof{figure}{7 variables 3-SAT problem converted to MaxSAT and codified as a graph}
    \label{fig:7_vars_sparse_problem}
\end{Figure}

The \textit{qpu\_sampling\_time}, $T_{s}$, is the time which determines the complexity. This time can be decomposed into:
\begin{equation}
T_{s}=\left(T_{a}+T_{r}+T_{d}\right)\cdot R
\end{equation}
Where $T_{a}$ is the single-sample annealing time, $T_{r}$ is the single-sample readout time and $T_{d}$ is the single-sample delay time, which consists of the following optional components:
\begin{equation}
T_{d}=readout\_thermalization+reduce\_intersample\_correlation+reinitialize\_state
\end{equation}

In a nutshell:
\begin{itemize}
    \item $T$: Total time of execution on the QPU.
    \item $T_{p}$: Total time to program the QPU. It depends on the $programming\_thermalization$, that is the time to wait after programming the QPU to cool back to base temperature. The value of the $programming\_thermalization$ parameter is the default one.
    \item $T_{s}$: Total time for all samples.
    \item $R$: Number of samples. In our case the number of samples equals the \textbf{number of variables in the QUBO formulation} of the problem.
    \item $T_{a}$: Annealing time per sample. In our case is the one by \textbf{default}.
    \item $T_{d}$: Delay between each pair of consecutive samples:
    \begin{itemize}
        \item $readout\_thermalization$: Time to wait after each measurement of a qubit for it to cool back to base temperature. In our case it is set to \textbf{$10\ \mu s$}.
        \item $reduce\_intersample\_correlation$: Delay time added after each anneal  for the sake of removing the effects from previous measurements of qubits.
        Adds a delay times before each anneal to lose the effects from the previous read. In our case it is set to \textbf{True} and it can be computed as:
        \begin{equation}
            delay=500+\frac{T_{schedule}\left(10000-500\right)}{2000}
        \end{equation}
        Where $T_{schedule}$ is the total time of the anneal schedule.
        \item $reinitialize\_state$: Time only used for reverse annealing. In our case it is set to \textbf{False}
    \end{itemize}
    \item $T_{r}$: Time per sample read. Depends on the number of qubits.
\end{itemize}

  


Fig. \ref{fig:quantum-linear} shows the average time taken for the execution of the testbed described in section \ref{subsec:annealing-circuit-testing-the-proposal}.
\begin{Figure}
    \centering
    \includegraphics[width=0.8\linewidth]{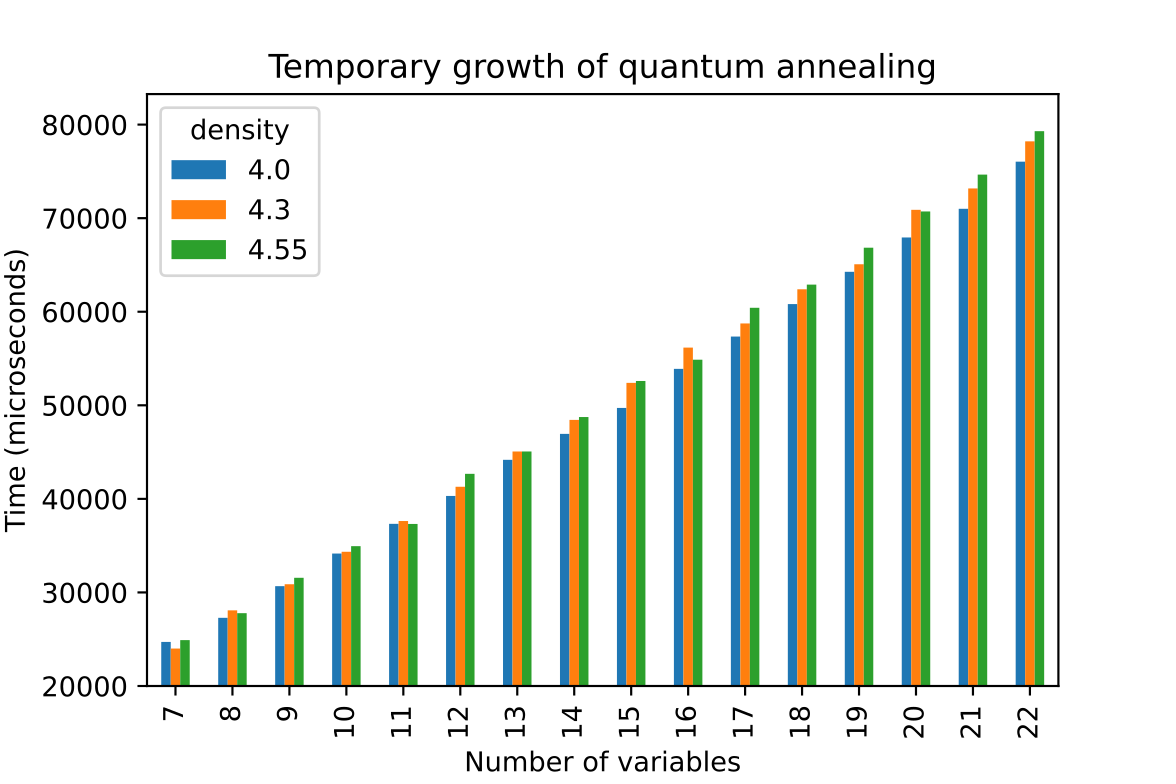}
    \captionof{figure}{Average QPU access time for testing}
    \label{fig:quantum-linear}
\end{Figure}

\section{Simulated annealing belongs to O(n)}\label{anexo:simulated-linear}
Each of the 5 executions of each of the 480 examples depends on three parameters, in a similar way to quantum annealing. These three parameters are:
\begin{itemize}
    \item \textit{num\_reads}: Number of executions of the simulated annealing algorithm. In our case it is set to the \textbf{number of
variables in the QUBO formulation} of the problem (same as for quantum annealing).
    \item \textit{num\_sweeps}: Number of steps in which the annealing protocol is divided. In our case it is set to \textbf{default} (same as for quantum annealing).
    \item \textit{num\_sweeps\_per\_beta}: Number of steps performed in each step of the annealing protocol. In our case it is set to \textbf{1}.
\end{itemize}
Thus, the complexity will depend on $O(num\_reads\cdot num\_sweeps\cdot num\_sweeps\_per\_beta)$.
Fig. \ref{fig:simulated-linear} shows the average time taken by the execution of the setups described in section \ref{subsec:annealing-circuit-testing-the-proposal}.
\begin{Figure}
    \centering
    \includegraphics[width=0.8\linewidth]{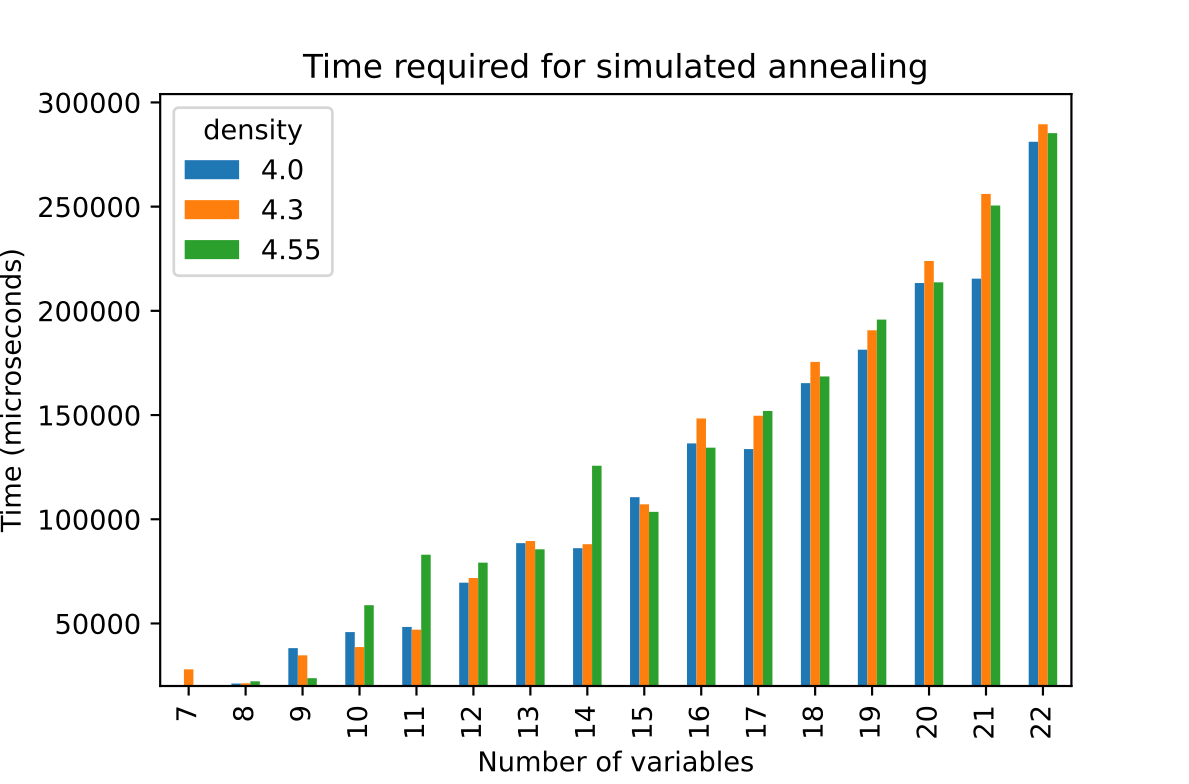}
    \captionof{figure}{Average execution time within our testbed}
    \label{fig:simulated-linear}
\end{Figure}

\newpage

\section{Accuracy of both annealing processes}\label{anexo:results-quantum-simulado-distance}
The following table compares Hamming and Cyclical distances between the state provided by each of the annealing options and, the solution state of the problem:

\begin{table}[h]\centering
\begin{tabular}{c|rr|rr|}
\cline{2-5}
\multicolumn{1}{l|}{}                                                                   & \multicolumn{2}{c|}{Hamming distance}                                                                                                                                   & \multicolumn{2}{c|}{Cyclical distance}                                                                                                                                   \\ \hline
\multicolumn{1}{|c|}{\begin{tabular}[c]{@{}c@{}}Number \\ of \\ variables\end{tabular}} & \multicolumn{1}{c|}{\begin{tabular}[c]{@{}c@{}}Quantum \\ annealing\end{tabular}} & \multicolumn{1}{c|}{\begin{tabular}[c]{@{}c@{}}Simulated \\ annealing\end{tabular}} & \multicolumn{1}{c|}{\begin{tabular}[c]{@{}c@{}}Quantum \\ annealing\end{tabular}} & \multicolumn{1}{c|}{\begin{tabular}[c]{@{}c@{}}Simulated \\ annealing\end{tabular}} \\ \hline
\multicolumn{1}{|c|}{7}     & \multicolumn{1}{r|}{1.33}      & 0.27  & \multicolumn{1}{r|}{13.73}  & 2.49     \\ \hline
\multicolumn{1}{|c|}{8}     & \multicolumn{1}{r|}{1.03}      & 0.19  & \multicolumn{1}{r|}{19.23}  & 7.23     \\ \hline
\multicolumn{1}{|c|}{9}     & \multicolumn{1}{r|}{1.9}      & 0.37  & \multicolumn{1}{r|}{73.5}  & 10.59     \\ \hline
\multicolumn{1}{|c|}{10}     & \multicolumn{1}{r|}{3.03}      & 0.47  & \multicolumn{1}{r|}{174.23}  & 24.35     \\ \hline
\multicolumn{1}{|c|}{11}     & \multicolumn{1}{r|}{3.83}      & 0.64  & \multicolumn{1}{r|}{357.6}  & 90.72     \\ \hline
\multicolumn{1}{|c|}{12}     & \multicolumn{1}{r|}{3.7}      & 1.02  & \multicolumn{1}{r|}{820.43}  & 189.55     \\ \hline
\multicolumn{1}{|c|}{13}     & \multicolumn{1}{r|}{4.17}      & 1.22  & \multicolumn{1}{r|}{1637.63}  & 494.97     \\ \hline
\multicolumn{1}{|c|}{14}     & \multicolumn{1}{r|}{4.67}      & 1.27  & \multicolumn{1}{r|}{3103.73}  & 936.16     \\ \hline
\multicolumn{1}{|c|}{15}     & \multicolumn{1}{r|}{4.93}      & 1.95  & \multicolumn{1}{r|}{6256.4}  & 2050.91     \\ \hline
\multicolumn{1}{|c|}{16}     & \multicolumn{1}{r|}{5.8}      & 2.81  & \multicolumn{1}{r|}{13125.23}  & 6471.44     \\ \hline
\multicolumn{1}{|c|}{17}     & \multicolumn{1}{r|}{7.03}      & 3.22  & \multicolumn{1}{r|}{30292.83}  & 12988.37     \\ \hline
\multicolumn{1}{|c|}{18}     & \multicolumn{1}{r|}{7.2}      & 3.81  & \multicolumn{1}{r|}{51982.17}  & 34910.17     \\ \hline
\multicolumn{1}{|c|}{19}     & \multicolumn{1}{r|}{7.93}      & 3.59  & \multicolumn{1}{r|}{134792.1}  & 49551.77     \\ \hline
\multicolumn{1}{|c|}{20}     & \multicolumn{1}{r|}{7.33}      & 4.82  & \multicolumn{1}{r|}{254491.07}  & 161184.32     \\ \hline
\multicolumn{1}{|c|}{21}     & \multicolumn{1}{r|}{8.53}      & 4.77  & \multicolumn{1}{r|}{576134.93}  & 260356.66     \\ \hline
\multicolumn{1}{|c|}{22}     & \multicolumn{1}{r|}{8.73}      & 5.25  & \multicolumn{1}{r|}{1013263.87}  & 563339.43     \\ \hline
\end{tabular}
\caption{Average distances of both annealing processes}
\end{table}

\newpage

\section{Quantum circuit-model search results}
\label{sec:anexo-model-circuit-results}

The following table compares Hamming and Cyclical quantum search proposed strategies with respect to original Grover's algorithm. Each of the strategies has been split into the case of knowing in advance the corresponding distance between annealing output state and actual solution state, and, not knowing that value in advance:

\begin{table}[h]\centering
\resizebox{\textwidth}{!}{
\begin{tabular}{cc|r|rr|rr|}
\cline{3-7}
                                                                                        &         & \multicolumn{1}{c|}{\multirow{2}{*}{\begin{tabular}[c]{@{}c@{}} \\ Grover's \\ algorithm\end{tabular}}} & \multicolumn{2}{c|}{\begin{tabular}[c]{@{}c@{}}Hamming  quantum search\end{tabular}}                                                                          & \multicolumn{2}{c|}{\begin{tabular}[c]{@{}c@{}}Cyclical  quantum  search\end{tabular}}                                                                         \\ \cline{1-2} \cline{4-7} 
\multicolumn{1}{|c|}{\begin{tabular}[c]{@{}c@{}}Number \\ of \\ variables\end{tabular}} & Density & \multicolumn{1}{c|}{}                                                                               & \multicolumn{1}{c|}{\begin{tabular}[c]{@{}c@{}}Unknown \\ distance\end{tabular}} & \multicolumn{1}{c|}{\begin{tabular}[c]{@{}c@{}}Known \\ distance\end{tabular}} & \multicolumn{1}{c|}{\begin{tabular}[c]{@{}c@{}}Unknown \\ distance\end{tabular}} & \multicolumn{1}{c|}{\begin{tabular}[c]{@{}c@{}}Known \\ distance\end{tabular}} \\ \hline
\multicolumn{1}{|c|}{7}  & 4.0 & 8.0 \ \ \  \textcolor{gray}{0.0\%}  & \multicolumn{1}{r|}{2.42 \textcolor{green}{ -69.75\%} } & 0.85 \textcolor{green}{ -89.42\%}  & \multicolumn{1}{r|}{1.32 \textcolor{green}{ -83.5\%} } & 0.78 \textcolor{green}{ -90.25\%}  \\ \hline
\multicolumn{1}{|c|}{7}  & 4.3 & 8.0 \ \ \  \textcolor{gray}{0.0\%}  & \multicolumn{1}{r|}{0.68 \textcolor{green}{ -91.5\%} } & 0.32 \textcolor{green}{ -95.97\%}  & \multicolumn{1}{r|}{0.24 \textcolor{green}{ -97.0\%} } & 0.24 \textcolor{green}{ -97.0\%}  \\ \hline
\multicolumn{1}{|c|}{7}  & 4.55 & 8.0 \ \ \  \textcolor{gray}{0.0\%}  & \multicolumn{1}{r|}{0.88 \textcolor{green}{ -89.0\%} } & 0.51 \textcolor{green}{ -93.59\%}  & \multicolumn{1}{r|}{0.48 \textcolor{green}{ -94.0\%} } & 0.44 \textcolor{green}{ -94.5\%}  \\ \hline
\multicolumn{1}{|c|}{8}  & 4.0 & 12.0 \ \ \  \textcolor{gray}{0.0\%}  & \multicolumn{1}{r|}{1.02 \textcolor{green}{ -91.5\%} } & 0.57 \textcolor{green}{ -95.22\%}  & \multicolumn{1}{r|}{1.12 \textcolor{green}{ -90.67\%} } & 0.6 \textcolor{green}{ -95.0\%}  \\ \hline
\multicolumn{1}{|c|}{8}  & 4.3 & 12.0 \ \ \  \textcolor{gray}{0.0\%}  & \multicolumn{1}{r|}{0.44 \textcolor{green}{ -96.33\%} } & 0.38 \textcolor{green}{ -96.83\%}  & \multicolumn{1}{r|}{1.12 \textcolor{green}{ -90.67\%} } & 1.0 \textcolor{green}{ -91.67\%}  \\ \hline
\multicolumn{1}{|c|}{8}  & 4.55 & 12.0 \ \ \  \textcolor{gray}{0.0\%}  & \multicolumn{1}{r|}{0.92 \textcolor{green}{ -92.33\%} } & 0.68 \textcolor{green}{ -94.32\%}  & \multicolumn{1}{r|}{1.92 \textcolor{green}{ -84.0\%} } & 1.52 \textcolor{green}{ -87.33\%}  \\ \hline
\multicolumn{1}{|c|}{9}  & 4.0 & 17.0 \ \ \  \textcolor{gray}{0.0\%}  & \multicolumn{1}{r|}{2.4 \textcolor{green}{ -85.88\%} } & 1.09 \textcolor{green}{ -93.58\%}  & \multicolumn{1}{r|}{1.68 \textcolor{green}{ -90.12\%} } & 1.06 \textcolor{green}{ -93.76\%}  \\ \hline
\multicolumn{1}{|c|}{9}  & 4.3 & 17.0 \ \ \  \textcolor{gray}{0.0\%}  & \multicolumn{1}{r|}{4.68 \textcolor{green}{ -72.47\%} } & 2.12 \textcolor{green}{ -87.51\%}  & \multicolumn{1}{r|}{2.64 \textcolor{green}{ -84.47\%} } & 1.64 \textcolor{green}{ -90.35\%}  \\ \hline
\multicolumn{1}{|c|}{9}  & 4.55 & 17.0 \ \ \  \textcolor{gray}{0.0\%}  & \multicolumn{1}{r|}{1.1 \textcolor{green}{ -93.53\%} } & 0.78 \textcolor{green}{ -95.41\%}  & \multicolumn{1}{r|}{1.68 \textcolor{green}{ -90.12\%} } & 0.78 \textcolor{green}{ -95.41\%}  \\ \hline
\multicolumn{1}{|c|}{10}  & 4.0 & 25.0 \ \ \  \textcolor{gray}{0.0\%}  & \multicolumn{1}{r|}{4.92 \textcolor{green}{ -80.32\%} } & 1.9 \textcolor{green}{ -92.39\%}  & \multicolumn{1}{r|}{3.74 \textcolor{green}{ -85.04\%} } & 2.74 \textcolor{green}{ -89.04\%}  \\ \hline
\multicolumn{1}{|c|}{10}  & 4.3 & 25.0 \ \ \  \textcolor{gray}{0.0\%}  & \multicolumn{1}{r|}{8.56 \textcolor{green}{ -65.76\%} } & 3.93 \textcolor{green}{ -84.29\%}  & \multicolumn{1}{r|}{4.08 \textcolor{green}{ -83.68\%} } & 2.84 \textcolor{green}{ -88.64\%}  \\ \hline
\multicolumn{1}{|c|}{10}  & 4.55 & 25.0 \ \ \  \textcolor{gray}{0.0\%}  & \multicolumn{1}{r|}{1.5 \textcolor{green}{ -94.0\%} } & 0.18 \textcolor{green}{ -99.27\%}  & \multicolumn{1}{r|}{0.34 \textcolor{green}{ -98.64\%} } & 0.34 \textcolor{green}{ -98.64\%}  \\ \hline
\multicolumn{1}{|c|}{11}  & 4.0 & 35.0 \ \ \  \textcolor{gray}{0.0\%}  & \multicolumn{1}{r|}{8.08 \textcolor{green}{ -76.91\%} } & 3.29 \textcolor{green}{ -90.6\%}  & \multicolumn{1}{r|}{8.0 \textcolor{green}{ -77.14\%} } & 5.06 \textcolor{green}{ -85.54\%}  \\ \hline
\multicolumn{1}{|c|}{11}  & 4.3 & 35.0 \ \ \  \textcolor{gray}{0.0\%}  & \multicolumn{1}{r|}{7.26 \textcolor{green}{ -79.26\%} } & 3.27 \textcolor{green}{ -90.65\%}  & \multicolumn{1}{r|}{8.5 \textcolor{green}{ -75.71\%} } & 6.1 \textcolor{green}{ -82.57\%}  \\ \hline
\multicolumn{1}{|c|}{11}  & 4.55 & 35.0 \ \ \  \textcolor{gray}{0.0\%}  & \multicolumn{1}{r|}{9.26 \textcolor{green}{ -73.54\%} } & 3.42 \textcolor{green}{ -90.22\%}  & \multicolumn{1}{r|}{5.5 \textcolor{green}{ -84.29\%} } & 3.32 \textcolor{green}{ -90.51\%}  \\ \hline
\multicolumn{1}{|c|}{12}  & 4.0 & 50.0 \ \ \  \textcolor{gray}{0.0\%}  & \multicolumn{1}{r|}{15.78 \textcolor{green}{ -68.44\%} } & 5.4 \textcolor{green}{ -89.2\%}  & \multicolumn{1}{r|}{9.8 \textcolor{green}{ -80.4\%} } & 6.64 \textcolor{green}{ -86.72\%}  \\ \hline
\multicolumn{1}{|c|}{12}  & 4.3 & 50.0 \ \ \  \textcolor{gray}{0.0\%}  & \multicolumn{1}{r|}{30.7 \textcolor{green}{ -38.6\%} } & 9.19 \textcolor{green}{ -81.63\%}  & \multicolumn{1}{r|}{14.7 \textcolor{green}{ -70.6\%} } & 9.8 \textcolor{green}{ -80.4\%}  \\ \hline
\multicolumn{1}{|c|}{12}  & 4.55 & 50.0 \ \ \  \textcolor{gray}{0.0\%}  & \multicolumn{1}{r|}{12.38 \textcolor{green}{ -75.24\%} } & 4.89 \textcolor{green}{ -90.23\%}  & \multicolumn{1}{r|}{9.1 \textcolor{green}{ -81.8\%} } & 5.5 \textcolor{green}{ -89.0\%}  \\ \hline
\multicolumn{1}{|c|}{13}  & 4.0 & 71.0 \ \ \  \textcolor{gray}{0.0\%}  & \multicolumn{1}{r|}{46.38 \textcolor{green}{ -34.68\%} } & 16.19 \textcolor{green}{ -77.2\%}  & \multicolumn{1}{r|}{32.0 \textcolor{green}{ -54.93\%} } & 20.46 \textcolor{green}{ -71.18\%}  \\ \hline
\multicolumn{1}{|c|}{13}  & 4.3 & 71.0 \ \ \  \textcolor{gray}{0.0\%}  & \multicolumn{1}{r|}{23.64 \textcolor{green}{ -66.7\%} } & 6.23 \textcolor{green}{ -91.23\%}  & \multicolumn{1}{r|}{17.0 \textcolor{green}{ -76.06\%} } & 9.98 \textcolor{green}{ -85.94\%}  \\ \hline
\multicolumn{1}{|c|}{13}  & 4.55 & 71.0 \ \ \  \textcolor{gray}{0.0\%}  & \multicolumn{1}{r|}{10.92 \textcolor{green}{ -84.62\%} } & 5.69 \textcolor{green}{ -91.98\%}  & \multicolumn{1}{r|}{18.0 \textcolor{green}{ -74.65\%} } & 7.34 \textcolor{green}{ -89.66\%}  \\ \hline
\multicolumn{1}{|c|}{14}  & 4.0 & 100.0 \ \ \  \textcolor{gray}{0.0\%}  & \multicolumn{1}{r|}{45.6 \textcolor{green}{ -54.4\%} } & 15.44 \textcolor{green}{ -84.56\%}  & \multicolumn{1}{r|}{25.56 \textcolor{green}{ -74.44\%} } & 17.06 \textcolor{green}{ -82.94\%}  \\ \hline
\multicolumn{1}{|c|}{14}  & 4.3 & 100.0 \ \ \  \textcolor{gray}{0.0\%}  & \multicolumn{1}{r|}{35.46 \textcolor{green}{ -64.54\%} } & 12.87 \textcolor{green}{ -87.13\%}  & \multicolumn{1}{r|}{32.66 \textcolor{green}{ -67.34\%} } & 22.34 \textcolor{green}{ -77.66\%}  \\ \hline
\multicolumn{1}{|c|}{14}  & 4.55 & 100.0 \ \ \  \textcolor{gray}{0.0\%}  & \multicolumn{1}{r|}{37.04 \textcolor{green}{ -62.96\%} } & 18.7 \textcolor{green}{ -81.3\%}  & \multicolumn{1}{r|}{22.72 \textcolor{green}{ -77.28\%} } & 13.86 \textcolor{green}{ -86.14\%}  \\ \hline

\end{tabular}}
\caption{Total iterations results for 7 to 14 variables}
\label{tab:total-iterations-7-14}
\end{table}

\begin{table}[h]\centering
\resizebox{\textwidth}{!}{
\begin{tabular}{cc|r|rr|rr|}
\cline{3-7}
                                                                                        &         & \multicolumn{1}{c|}{\multirow{2}{*}{\begin{tabular}[c]{@{}c@{}} \\ Grover's \\ algorithm\end{tabular}}} & \multicolumn{2}{c|}{\begin{tabular}[c]{@{}c@{}}Hamming  quantum search\end{tabular}}                                                                          & \multicolumn{2}{c|}{\begin{tabular}[c]{@{}c@{}}Cyclical  quantum  search\end{tabular}}                                                                         \\ \cline{1-2} \cline{4-7} 
\multicolumn{1}{|c|}{\begin{tabular}[c]{@{}c@{}}Number \\ of \\ variables\end{tabular}} & Density & \multicolumn{1}{c|}{}                                                                               & \multicolumn{1}{c|}{\begin{tabular}[c]{@{}c@{}}Unknown \\ distance\end{tabular}} & \multicolumn{1}{c|}{\begin{tabular}[c]{@{}c@{}}Known \\ distance\end{tabular}} & \multicolumn{1}{c|}{\begin{tabular}[c]{@{}c@{}}Unknown \\ distance\end{tabular}} & \multicolumn{1}{c|}{\begin{tabular}[c]{@{}c@{}}Known \\ distance\end{tabular}} \\ \hline
\multicolumn{1}{|c|}{15}  & 4.0 & 142.0 \ \ \  \textcolor{gray}{0.0\%}  & \multicolumn{1}{r|}{126.94 \textcolor{green}{ -10.61\%} } & 38.74 \textcolor{green}{ -72.72\%}  & \multicolumn{1}{r|}{70.0 \textcolor{green}{ -50.7\%} } & 45.42 \textcolor{green}{ -68.01\%}  \\ \hline
\multicolumn{1}{|c|}{15}  & 4.3 & 142.0 \ \ \  \textcolor{gray}{0.0\%}  & \multicolumn{1}{r|}{79.02 \textcolor{green}{ -44.35\%} } & 25.84 \textcolor{green}{ -81.8\%}  & \multicolumn{1}{r|}{44.0 \textcolor{green}{ -69.01\%} } & 27.34 \textcolor{green}{ -80.75\%}  \\ \hline
\multicolumn{1}{|c|}{15}  & 4.55 & 142.0 \ \ \  \textcolor{gray}{0.0\%}  & \multicolumn{1}{r|}{60.2 \textcolor{green}{ -57.61\%} } & 22.2 \textcolor{green}{ -84.36\%}  & \multicolumn{1}{r|}{28.0 \textcolor{green}{ -80.28\%} } & 18.42 \textcolor{green}{ -87.03\%}  \\ \hline
\multicolumn{1}{|c|}{16}  & 4.0 & 201.0 \ \ \  \textcolor{gray}{0.0\%}  & \multicolumn{1}{r|}{144.36 \textcolor{green}{ -28.18\%} } & 44.04 \textcolor{green}{ -78.09\%}  & \multicolumn{1}{r|}{79.52 \textcolor{green}{ -60.44\%} } & 57.68 \textcolor{green}{ -71.3\%}  \\ \hline
\multicolumn{1}{|c|}{16}  & 4.3 & 201.0 \ \ \  \textcolor{gray}{0.0\%}  & \multicolumn{1}{r|}{143.98 \textcolor{green}{ -28.37\%} } & 41.69 \textcolor{green}{ -79.26\%}  & \multicolumn{1}{r|}{88.04 \textcolor{green}{ -56.2\%} } & 57.46 \textcolor{green}{ -71.41\%}  \\ \hline
\multicolumn{1}{|c|}{16}  & 4.55 & 201.0 \ \ \  \textcolor{gray}{0.0\%}  & \multicolumn{1}{r|}{240.12 \textcolor{red}{ +19.46\%} } & 64.23 \textcolor{green}{ -68.05\%}  & \multicolumn{1}{r|}{99.4 \textcolor{green}{ -50.55\%} } & 74.68 \textcolor{green}{ -62.85\%}  \\ \hline
\multicolumn{1}{|c|}{17}  & 4.0 & 284.0 \ \ \  \textcolor{gray}{0.0\%}  & \multicolumn{1}{r|}{417.28 \textcolor{red}{ +46.93\%} } & 136.04 \textcolor{green}{ -52.1\%}  & \multicolumn{1}{r|}{225.12 \textcolor{green}{ -20.73\%} } & 154.1 \textcolor{green}{ -45.74\%}  \\ \hline
\multicolumn{1}{|c|}{17}  & 4.3 & 284.0 \ \ \  \textcolor{gray}{0.0\%}  & \multicolumn{1}{r|}{179.84 \textcolor{green}{ -36.68\%} } & 65.99 \textcolor{green}{ -76.76\%}  & \multicolumn{1}{r|}{124.62 \textcolor{green}{ -56.12\%} } & 59.74 \textcolor{green}{ -78.96\%}  \\ \hline
\multicolumn{1}{|c|}{17}  & 4.55 & 284.0 \ \ \  \textcolor{gray}{0.0\%}  & \multicolumn{1}{r|}{128.36 \textcolor{green}{ -54.8\%} } & 49.54 \textcolor{green}{ -82.56\%}  & \multicolumn{1}{r|}{124.62 \textcolor{green}{ -56.12\%} } & 71.5 \textcolor{green}{ -74.82\%}  \\ \hline
\multicolumn{1}{|c|}{18}  & 4.0 & 402.0 \ \ \  \textcolor{gray}{0.0\%}  & \multicolumn{1}{r|}{415.18 \textcolor{red}{ +3.28\%} } & 140.15 \textcolor{green}{ -65.14\%}  & \multicolumn{1}{r|}{238.56 \textcolor{green}{ -40.66\%} } & 165.06 \textcolor{green}{ -58.94\%}  \\ \hline
\multicolumn{1}{|c|}{18}  & 4.3 & 402.0 \ \ \  \textcolor{gray}{0.0\%}  & \multicolumn{1}{r|}{377.02 \textcolor{green}{ -6.21\%} } & 107.89 \textcolor{green}{ -73.16\%}  & \multicolumn{1}{r|}{210.16 \textcolor{green}{ -47.72\%} } & 131.36 \textcolor{green}{ -67.32\%}  \\ \hline
\multicolumn{1}{|c|}{18}  & 4.55 & 402.0 \ \ \  \textcolor{gray}{0.0\%}  & \multicolumn{1}{r|}{453.74 \textcolor{red}{ +12.87\%} } & 158.91 \textcolor{green}{ -60.47\%}  & \multicolumn{1}{r|}{266.96 \textcolor{green}{ -33.59\%} } & 188.64 \textcolor{green}{ -53.07\%}  \\ \hline
\multicolumn{1}{|c|}{19}  & 4.0 & 568.0 \ \ \  \textcolor{gray}{0.0\%}  & \multicolumn{1}{r|}{541.06 \textcolor{green}{ -4.74\%} } & 159.3 \textcolor{green}{ -71.95\%}  & \multicolumn{1}{r|}{393.96 \textcolor{green}{ -30.64\%} } & 225.36 \textcolor{green}{ -60.32\%}  \\ \hline
\multicolumn{1}{|c|}{19}  & 4.3 & 568.0 \ \ \  \textcolor{gray}{0.0\%}  & \multicolumn{1}{r|}{348.6 \textcolor{green}{ -38.63\%} } & 118.28 \textcolor{green}{ -79.18\%}  & \multicolumn{1}{r|}{281.4 \textcolor{green}{ -50.46\%} } & 177.86 \textcolor{green}{ -68.69\%}  \\ \hline
\multicolumn{1}{|c|}{19}  & 4.55 & 568.0 \ \ \  \textcolor{gray}{0.0\%}  & \multicolumn{1}{r|}{470.74 \textcolor{green}{ -17.12\%} } & 170.67 \textcolor{green}{ -69.95\%}  & \multicolumn{1}{r|}{257.28 \textcolor{green}{ -54.7\%} } & 156.56 \textcolor{green}{ -72.44\%}  \\ \hline
\multicolumn{1}{|c|}{20}  & 4.0 & 804.0 \ \ \  \textcolor{gray}{0.0\%}  & \multicolumn{1}{r|}{1156.86 \textcolor{red}{ +43.89\%} } & 337.54 \textcolor{green}{ -58.02\%}  & \multicolumn{1}{r|}{624.8 \textcolor{green}{ -22.29\%} } & 405.58 \textcolor{green}{ -49.55\%}  \\ \hline
\multicolumn{1}{|c|}{20}  & 4.3 & 804.0 \ \ \  \textcolor{gray}{0.0\%}  & \multicolumn{1}{r|}{749.66 \textcolor{green}{ -6.76\%} } & 258.34 \textcolor{green}{ -67.87\%}  & \multicolumn{1}{r|}{590.72 \textcolor{green}{ -26.53\%} } & 410.28 \textcolor{green}{ -48.97\%}  \\ \hline
\multicolumn{1}{|c|}{20}  & 4.55 & 804.0 \ \ \  \textcolor{gray}{0.0\%}  & \multicolumn{1}{r|}{923.16 \textcolor{red}{ +14.82\%} } & 303.85 \textcolor{green}{ -62.21\%}  & \multicolumn{1}{r|}{477.12 \textcolor{green}{ -40.66\%} } & 330.5 \textcolor{green}{ -58.89\%}  \\ \hline
\multicolumn{1}{|c|}{21}  & 4.0 & 1137.0 \ \ \  \textcolor{gray}{0.0\%}  & \multicolumn{1}{r|}{1429.72 \textcolor{red}{ +25.74\%} } & 412.86 \textcolor{green}{ -63.69\%}  & \multicolumn{1}{r|}{755.76 \textcolor{green}{ -33.53\%} } & 528.68 \textcolor{green}{ -53.5\%}  \\ \hline
\multicolumn{1}{|c|}{21}  & 4.3 & 1137.0 \ \ \  \textcolor{gray}{0.0\%}  & \multicolumn{1}{r|}{662.64 \textcolor{green}{ -41.72\%} } & 276.12 \textcolor{green}{ -75.72\%}  & \multicolumn{1}{r|}{755.76 \textcolor{green}{ -33.53\%} } & 456.82 \textcolor{green}{ -59.82\%}  \\ \hline
\multicolumn{1}{|c|}{21}  & 4.55 & 1137.0 \ \ \  \textcolor{gray}{0.0\%}  & \multicolumn{1}{r|}{1297.7 \textcolor{red}{ +14.13\%} } & 379.97 \textcolor{green}{ -66.58\%}  & \multicolumn{1}{r|}{836.16 \textcolor{green}{ -26.46\%} } & 547.52 \textcolor{green}{ -51.85\%}  \\ \hline
\multicolumn{1}{|c|}{22}  & 4.0 & 1608.0 \ \ \  \textcolor{gray}{0.0\%}  & \multicolumn{1}{r|}{2036.14 \textcolor{red}{ +26.63\%} } & 633.04 \textcolor{green}{ -60.63\%}  & \multicolumn{1}{r|}{932.34 \textcolor{green}{ -42.02\%} } & 604.42 \textcolor{green}{ -62.41\%}  \\ \hline
\multicolumn{1}{|c|}{22}  & 4.3 & 1608.0 \ \ \  \textcolor{gray}{0.0\%}  & \multicolumn{1}{r|}{1947.88 \textcolor{red}{ +21.14\%} } & 638.87 \textcolor{green}{ -60.27\%}  & \multicolumn{1}{r|}{1273.44 \textcolor{green}{ -20.81\%} } & 714.5 \textcolor{green}{ -55.57\%}  \\ \hline
\multicolumn{1}{|c|}{22}  & 4.55 & 1608.0 \ \ \  \textcolor{gray}{0.0\%}  & \multicolumn{1}{r|}{1516.18 \textcolor{green}{ -5.71\%} } & 559.94 \textcolor{green}{ -65.18\%}  & \multicolumn{1}{r|}{1182.48 \textcolor{green}{ -26.46\%} } & 720.12 \textcolor{green}{ -55.22\%}  \\ \hline

\end{tabular}}
\caption{Total iterations results for 15 to 22 variables}
\label{tab:total-iterations-15-22}
\end{table}



\end{document}